\documentclass[aps,prb,twocolumn, superscriptaddress, showpacs]{revtex4-2}
\usepackage{amsmath,amssymb,amsfonts}

\usepackage{comment}
\usepackage{scalerel}
\usepackage{mathtools}
\usepackage{physics}
\usepackage{float}
\usepackage{graphicx}
\usepackage{dcolumn}
\usepackage{bm}
\usepackage{hyperref}
\hypersetup{colorlinks, linkcolor={red}, citecolor = {blue}, urlcolor={blue}}

\usepackage{xcolor}
\usepackage{soul}
\usepackage{empheq}

\let\oldref\ref
\renewcommand{\ref}[1]{(\oldref{#1})}

\def\CFT{\textsl{\tiny CFT}}

\newcommand{\J}[2]{J_{#1}^{#2, \mbox{\textsl{\tiny CFT}}}} 
\newcommand{\Jbar}[2]{\bar{J}_{#1}^{#2, \mbox{\textsl{\tiny  CFT}}}}

\begin{document}


\title{Emergence of Kac--Moody Symmetry in Critical Quantum Spin Chains}

\author{Ruoshui Wang}
\email{rw552@cornell.edu}
\affiliation{Cornell University, Ithaca, NY 14853 USA}

\author{Yijian Zou}
\affiliation{Stanford Institute for Theoretical Physics, Stanford University, Palo Alto, CA 94305, USA}

\author{Guifr\'e Vidal}
\affiliation{Google Quantum AI, Mountain View, CA 94043, USA}

\newcommand {\hide} [1] {}

\date{\today}

\begin{abstract}

Given a critical quantum spin chain with a microscopic Lie-group symmetry, corresponding e.g. to $U(1)$ or $SU(2)$ spin isotropy, we numerically investigate the emergence of Kac--Moody symmetry at low energies and long distances. In that regime, one such critical quantum spin chain is described by a conformal field theory where the usual Virasoro algebra associated to conformal invariance is augmented with a Kac--Moody algebra associated to conserved currents. Specifically, we first propose a method to construct lattice operators corresponding to the Kac--Moody generators. We then numerically show that, when projected onto low energy states of the quantum spin chain, these operators indeed approximately fulfill the Kac--Moody algebra. The lattice version of the Kac--Moody generators allow us to compute the so-called level constant and to organize the low-energy eigenstates of the lattice Hamiltonian into Kac--Moody towers. We illustrate the proposal with the XXZ model and the Heisenberg model with a next-to-nearest-neighbor coupling. 

\end{abstract}

\maketitle


\section{\label{sec:level1}Introduction}
Near a continuous phase transition, microscopically different systems may display low-energy, long-distance behaviour that is remarkably similar. 
Such universality is explained by saying that those systems flow to the same fixed point of the renormalization group (RG) \cite{wilson_renormalization_1974}. In 1+1 dimensions, the universality class of a (massless) RG fixed point is often described by a conformal field theory (CFT) which, in turn, is completely characterized by a set of parameters known as the conformal data \cite{Friedan_1984,belavin_infinite_1984}. 

In this work we are concerned with critical quantum spin chains, as specified in terms of a local lattice Hamiltonian in one spatial dimension. Given one such Hamiltonian, a natural goal is to numerically compute the conformal data of the emergent 1+1 CFT that describes the behaviour of the spin chain at long distances and low energies. In other words, to numerically characterize the emergent universal behaviour of the phase transition realized by the quantum spin chain.
For this purpose, one may follow an approach, initiated in the 1980's \cite{cardy1984,blote_1986,affleck1986,cardy1986}, that is based on the \textit{operator-state correspondence}. Given a 1+1 CFT on the circle, this correspondence establishes that each simultaneous eigenstate $|\psi^{\CFT}_\alpha\rangle$ of the Hamiltonian $H^{\CFT}$ and momentum $P^{\CFT}$ of the CFT on the circle corresponds to a scaling operator $\psi^{\CFT}_\alpha$ (an operator that transforms covariantly under scale transformations). Moreover, the energy and momentum of the state $|\psi^{\CFT}_\alpha\rangle$ relate to the universal scaling properties of the operator $\psi^{\CFT}_\alpha$. Importantly, the low energy states of a critical quantum spin chain on the circle are organized as in the emergent CFT. Accordingly, Cardy and others \cite{cardy1984,blote_1986,affleck1986,cardy1986} proposed that one could extract the conformal data by simply studying the low energy states of the critical quantum spin chain. Over the years, several other authors have contributed additional insights into this strategy. A crucial step was to identify operators on the lattice with operators of the CFT. In particular, certain lattice operators can be identified with Virasoro generators of the conformal symmetry in the CFT \cite{ks, READ2007316, DUBAIL2010399, Gainutdinov_2013, GAINUTDINOV2013223, Bondesan:2014hza, mv}. This means that such lattice operators act on the low-energy states of the critical quantum spin chain approximately (up to finite size corrections) in the same way as the Virasoro generators act on the corresponding states in the CFT. The lattice Virasoro generators allow us to see the emergence of conformal symmetry directly on the lattice, offering a way to systematically identify primary operators and conformal towers in the low-energy spectrum of a quantum critical spin chain \cite{mv, zmv1, zmv2}.

In some CFTs, conformal symmetry is enhanced to a larger symmetry, which further relates different Virasoro primary operators and conformal towers. Extended symmetries exist in many CFTs that describe lattice models, such as quantum spin chains \cite{Rahmani_2015,OBF_2018}, classical statistical-mechanics models \cite{FATEEV1987644,Deguchi2001}, and edge modes of topological orders \cite{Halperin_1982,Wen_1990}. One important implication of the extended symmetry is that it reduces the number of independent primary operators in the conformal data. In the cases known as rational CFTs \cite{moore1989classical}, the number of primary operators with respect to the extended symmetry is finite, while the number of Virasoro primary fields can be infinite. This makes the extraction of complete conformal data possible. A prominent example is when conformal symmetry is enhanced by a global symmetry with Lie group $G$. In this case, scaling operators are organized by an extension of the Virasoro algebra, the Kac--Moody algebra \cite{francesco2012,blumenhagen2009}, denoted by $\mathfrak{g}_k$, where $\mathfrak{g}$ is the Lie algebra of the the group $G$ and $k\in \mathbb{Z}$ is the \textit{level constant}. A remarkable consequence of the extension is that the global symmetry of the CFT becomes $G\times G$, which acts independently on left and right moving fields. 

In this paper we investigate how the Kac--Moody symmetry emerges from critical quantum spin chains with a Lie group symmetry $G$. We construct lattice operators that correspond to the generators of the Kac--Moody algebra. We numerically confirm that the global symmetry is enhanced to $G\times G$ in the low-energy subspace and that the eigenstates can be organized into Kac--Moody towers. We test our proposal with the XXZ model for the $G=U(1)$ case and the Heisenberg model with next-to-nearest-neighbor interactions for the $G=SU(2)$ case. In both cases we find that the proposed lattice Kac--Moody generators act on the low-energy states as their CFT counterparts in the thermodynamic limit. In particular, we can extract the level constant $k$ by computing the commutators of lattice Kac--Moody generators in the low-energy subspace. In previous work by some of the authors, lattice operators that correspond to generators of another form of extended symmetry, namely superconformal symmetry \cite{zv}, were found numerically. Here, in contrast, we provide an analytical ansatz for the lattice Kac--Moody generators. For the specific case of the XXZ model, which is an integrable model, the lattice Kac--Moody generators can also be found via bosonization \cite{fradkin2013} (as reviewed in Appendix~\ref{appendix:a}), and we can use this previous result to further validate our proposal. In contrast, no previous results appear to be known for the Heisenberg model with next-to-near-neighbor interactions. The performance of our method, which treats all models on the same footing, is seen to not rely on integrablity. 

The rest of the paper is organized as follows. In Sec.~\ref{sec:review}, we review the conformal symmetry and the Virasoro algebra in a CFT, as well as the approximate lattice version of Virasoro generators. In Sec.~\ref{sec:u1}, we consider the Kac--Moody algebra for the Abelian group $G=U(1)$. We first review the free boson CFT and the action of the Kac--Moody algebra, and then propose our method to build an approximate lattice version of the Kac--Moody generators. We also present numerical tests of the lattice Kac--Moody generators in the XXZ quantum spin model. In Sec.~\ref{sec:su2} we generalize the formalism to general non-Abelian group $G$. We will focus on $G=SU(2)$ and construct an approximate lattice version of the Kac--Moody generators for the Heisenberg model with next-to-nearest-neighbor interactions. In Sec.~\ref{sec:discussion} we conclude with discussions and future directions.
\section{Conformal symmetry and lattice Virasoro generators}
\label{sec:review}
In this section we review conformal symmetry, the Virasoro algebra and its approximate realization in the low-energy states of a quantum critical spin chain, see Ref.~\cite{mv} and Ref.~\cite{francesco2012} for an introduction. Throughout this paper, we use superscript $^\CFT$ to denote objects in a CFT, e.g. $T^\CFT$, $L_n^\CFT$, $\J{n}{\alpha}$, and in this way distinguish them from the corresponding objects in a lattice, e.g. $T$, $L_n$, $J_n^{\alpha}$.
\subsection{Virasoro algebra}
Consider a conformal field theory in 1+1 dimensions. Conformal transformations on the plane are generated by Virasoro generators $L^{\CFT}_n,\bar{L}^{\CFT}_n~(n\in\mathbb{Z})$ that satisfy the Virasoro algebra,
\begin{align}
\label{VirasoroCFT}
&[L^{\CFT}_n,L^{\CFT}_m]=(n-m) L^{\CFT}_{n+m} + \frac{c^{\CFT}}{12}n (n^2-1) \delta_{n+m,0} \\
&[\bar{L}^{\CFT}_n,\bar{L}^{\CFT}_m]=(n-m) \bar{L}^{\CFT}_{n+m} + \frac{c^{\CFT}}{12}n (n^2-1) \delta_{n+m,0} \\
&[L^{\CFT}_n,\bar{L}^{\CFT}_m]=0.
\end{align}
In particular, dilations and rotations of the plane are generated by $D^{\CFT}=L^{\CFT}_0+\bar{L}^{\CFT}_0$ and $R^{\CFT}=L^{\CFT}_0-\bar{L}^{\CFT}_0$, respectively. The Hilbert space of the CFT is supported on circles around the origin. Any state in the Hilbert space can be spanned by simultaneous eigenstates $|\psi^{\CFT}_\alpha\rangle$ of $L^{\CFT}_0$ and $\bar{L}^{\CFT}_0$,
\begin{equation}
    L^{\CFT}_0|\psi^{\CFT}_\alpha\rangle=h^{\CFT}_\alpha|\psi^{\CFT}_\alpha\rangle, ~~ \bar{L}^{\CFT}_0|\psi^{\CFT}_\alpha\rangle=\bar{h}^{\CFT}_\alpha|\psi^{\CFT}_\alpha\rangle,
\end{equation}
where $(h^{\CFT}_\alpha,\bar{h}^{\CFT}_\alpha)$ are the holomorphic and anti-holomorphic conformal dimensions of the state. The eigenvalues of $D^{\CFT}$ and $R^{\CFT}$ are $\Delta^{\CFT}_\alpha=h^{\CFT}_\alpha+\bar{h}^{\CFT}_\alpha$ and $s^{\CFT}_\alpha=h^{\CFT}_\alpha-\bar{h}^{\CFT}_\alpha$, known as scaling dimensions and conformal spins, respectively.

Setting $n=0$ in Eq.~\eqref{VirasoroCFT}, we see that acting with $L^{\CFT}_m$ on a state changes the holomorphic dimension by $-m$. Therefore, $L^{\CFT}_m$ lowers the scaling dimensions and the conformal spins by $m$. Similarly, $\bar{L}^{\CFT}_m$ lowers the scaling dimensions by $m$ and increases the conformal spins by $m$. A Virasoro primary state $|\phi^{\CFT}_\alpha\rangle$ is defined such that its scaling dimension cannot be lowered,
\begin{equation}
\label{eq:primarydef}
    L^{\CFT}_n|\phi^{\CFT}_\alpha\rangle=0, ~~ \bar{L}^{\CFT}_n|\phi^{\CFT}_\alpha\rangle=0, ~~ \forall n>0.
\end{equation}
Each primary state is associated with a Virasoro conformal tower, which contains that primary state as well as all its Virasoro descendant states. The descendant states are obtained by acting successively with raising operators $L^{\CFT}_m,\bar{L}^{\CFT}_m~(m<0)$ on the primary state.

\subsection{CFT on the cylinder}
In 1+1 dimensions one can use a conformal transformation to map the CFT from the plane to a cylinder $S^{1}\times\mathbb R$, where the axial direction is the imaginary time direction with coordinate $\tau\in(-\infty,\infty)$, and the angular direction is the spatial direction with coordinate $x\in [0,L)$. The Hilbert space is supported on the equal-time slices. We will focus on the $\tau=0$ slice, and denote the fields with its spatial coordinate $x$. In any CFT, there exist the holomorphic and anti-holomorphic energy-momentum tensors $T^\CFT$ and $\bar{T}^\CFT$, with conformal dimensions $(2,0)$ and $(0,2)$, respectively. The Hamiltonian and momentum can be expressed as an integral of the stress tensor,
\begin{eqnarray}
H^{\CFT} &=& \int_0^{L} dx\, h^\CFT(x) \\
P^{\CFT} &=& \int_0^{L} dx\, p^\CFT(x),
\end{eqnarray}
where
\begin{eqnarray}
h^\CFT(x) &=& \frac{1}{2\pi}(T^\CFT(x)+\bar{T}^\CFT(x)) \\
p^\CFT(x) &=& \frac{1}{2\pi}(T^\CFT(x)-\bar{T}^\CFT(x)).
\end{eqnarray}
The Virasoro generators are the Fourier modes of $T^\CFT$ and $\bar{T}^\CFT$,
\begin{equation}
\begin{aligned}
\label{eq:LnCFT}
L^{\CFT}_n = \frac{L}{(2\pi)^2} \int_{0}^{L} dx e^{+ i n x \frac{2\pi}{L}} T^\CFT (x) + \frac{c^{\CFT}}{24} \delta_{n,0} , \\
\bar{L}^{\CFT}_n = \frac{L}{(2\pi)^2} \int_{0}^{L} dx e^{- i n x \frac{2\pi}{L}} \bar{T}^\CFT (x) + \frac{c^{\CFT}}{24} \delta_{n,0} . 
\end{aligned}
\end{equation}
We may express the Hamiltonian and momentum as
\begin{eqnarray}
\label{HCFT}
H^{\CFT}&=&\frac{2\pi}{L}\left(L^{\CFT}_0+\bar{L}^{\CFT}_0-\frac{c^{\CFT}}{12}\right)  \\
\label{PCFT}
P^{\CFT}&=& \frac{2\pi}{L}\left(L^{\CFT}_0-\bar{L}^{\CFT}_0\right).
\end{eqnarray}
We see that each simultaneous eigenstate $|\psi^{\CFT}_\alpha\rangle$ of $H^{\CFT}$ and $P^{\CFT}$ has energy and momentum
\begin{eqnarray}
\label{eq:ECFT}
E^{\CFT}_\alpha&=&\frac{2\pi}{L}\left(\Delta^{\CFT}_\alpha-\frac{c^{\CFT}}{12}\right)  \\
\label{eq:PCFT}
P^{\CFT}_\alpha&=& \frac{2\pi}{L} s^{\CFT}_\alpha.
\end{eqnarray}
It is also useful to express the Fourier mode of the Hamiltonian density as a linear combination of Virasoro generators $L_n^{\CFT}$ and $\bar{L}_{-n}^{\CFT}$,
\begin{equation}
 L^{\CFT}_n+\bar{L}^{\CFT}_{-n}= H^{\CFT}_n \equiv \frac{L}{2\pi} \int_{0}^{L} dx e^{+ i n x \frac{2\pi}{L}} h^\CFT (x).
\end{equation}
These operators transform states within the same conformal tower of the CFT.
\subsection{Lattice Virasoro generators}

Given a critical quantum spin chain on a circle with Hamiltonian
\begin{equation}
    H=\sum_{j} h_j,
\end{equation}
its low-energy eigenstates $|\psi_\alpha\rangle$ are in one-to-one correspondence with CFT states $|\psi^{\CFT}_\alpha\rangle$. The energies and momenta of $|\psi_\alpha\rangle$ are related to the scaling dimensions and conformal spins by \cite{cardy1984,blote_1986,affleck1986}
\begin{eqnarray}
\label{eq:Elat}
E_\alpha &=& A+B\frac{2\pi}{N}\left(\Delta^{\CFT}_\alpha-\frac{c^{\CFT}}{12}\right)+O(N^{-x}) \\
\label{eq:Plat}
P_\alpha &=& \frac{2\pi}{N}s^{\CFT}_\alpha,
\end{eqnarray}
where the $O(N^{-x})$ with $x>1$ is the non-universal finite-size correction. These relations are direct lattice versions of expressions Eqs.~\eqref{eq:ECFT}-\eqref{eq:PCFT} for the CFT, in that upon identifying the circle size $L$ with the size $N$ of the spin chain, we could rewrite them as $E_{\alpha} = A + BE_{\alpha} + O(N^{-x})$ and $P_{\alpha} = P_{\alpha}^{\CFT}$. From Eqs.~\eqref{eq:Elat}-\eqref{eq:Plat}, one may extract approximate scaling dimensions $\Delta_\alpha$ and exact conformal spins $s_\alpha$ from the low-energy spectrum of the spin chain. In particular, the constants $A,B$ (which depend on how the lattice Hamiltonian is normalized) may be determined by demanding that the scaling dimensions of the identity operator and the stress tensor be $\Delta_{\mathbf{1}}=\Delta^{\CFT}_{\mathbf{1}}=0$ and $\Delta_{T}=\Delta^{\CFT}_T=2$ \cite{mv}. Notice that the states $\ket{\mathbf{1}^{\CFT}}$ and $\ket{T^{\CFT}}$ are present in any CFT, so the above procedure to determine constants $A$ and $B$ can always be applied.

The lattice operator $H_n$ that corresponds to $H^{\CFT}_n$ is the Fourier mode of the lattice Hamiltonian density $h_j$,
\begin{equation}
    H_n=\frac{N}{B}\sum_{j} e^{inj \frac{2\pi}{N}} h_j \sim H^{\CFT}_n.
\end{equation}
There is both analytical and numerical evidence that $H_n$ acts on the low-energy eigenstates of the lattice Hamiltonian as $H^{\CFT}_n$ acts on the corresponding CFT states \cite{ks, READ2007316, DUBAIL2010399, Gainutdinov_2013, GAINUTDINOV2013223, Bondesan:2014hza, mv}. We stress that the lattice Virasoro generators $H_n$ only (approximately) satisfy the algebra obeyed by the CFT operators $H^{\CFT}_n$ when projected onto the low-energy eigenstates of the quantum spin chains. At the level of lattice operators, it is easily checked (e.g. numerically) that the $H_n$ operators do not satisfy the Virasoro algebra. Similar to the $O(N^{-x})$ term in Eq.~\eqref{eq:Elat}, there are also non-universal finite-size corrections in the matrix elements of $H_n$, which can be reduced by an extrapolation to the thermodynamic limit (that is, the limit of an inifnitely large spin chain, $N \rightarrow \infty$).

Since the lattice Virasoro generators $H_n$ connect low-energy eigenstates within the same conformal tower, they can be used to identify Virasoro primary states and their conformal towers on the lattice \cite{mv}. 

\section{Lattice realization of Kac--Moody algebra: the $U(1)$ case}
\label{sec:u1}
In this section we first review the $U(1)$ Kac--Moody algebra and its manifestation in the free compactified boson CFT. We then consider a critical quantum spin chain with $U(1)$ symmetry and construct a lattice version of the Kac--Moody generators. Finally we numerically verify the actions of the proposed lattice Kac--Moody generators in the XXZ model. 

\subsection{$U(1)$ Kac--Moody algebra}
We consider a CFT with a global $U(1)$ symmetry, with conserved $U(1)$ charge $Q^{\CFT}$, which is an integral over space of a conserved local current $q^{\CFT}(x)$,
\begin{equation}
\label{eq:U1QCFT}
    Q^{\CFT}=\int dx\, q^{\CFT}(x).
\end{equation}
In a CFT, it turns out that the conserved current can be further divided into holomorphic and anti-holomorphic parts which are separately conserved,
\begin{equation}
\label{eq:U1qCFT}
    q^{\CFT}(x)=J^{\CFT}(x)+\bar{J}^{\CFT}(x),
\end{equation}
where $J^{\CFT}$ and $\bar{J}^{\CFT}$ are \textit{current operators} with conformal dimensions $(1,0)$ and $(0,1)$, respectively. Since both current operators satisfy conservation laws, namely $\bar{\partial} J^{\CFT}=0$ and $\partial \bar{J}^{\CFT}=0 $, there is an additional conserved charge
\begin{equation}
\label{eq:U1MCFT}
    M^{\CFT}=\int dx\, m^{\CFT}(x),
\end{equation}
where
\begin{equation}
\label{eq:U1mCFT}
    m^{\CFT}(x)=J^{\CFT}(x)-\bar{J}^{\CFT}(x).
\end{equation}
The global symmetry is therefore $U(1)\times U(1)$. 

For later reference, let us combine Eqs.~\eqref{eq:U1qCFT}-\eqref{eq:U1mCFT} and the conservation relations above, to obtain
\begin{equation}
    -i\partial_\tau q^{\CFT}(x)=\partial_x m^{\CFT}(x).
\end{equation}
Then, upon substituting the time derivative with the commutator by the Hamiltonian we further obtain 
\begin{equation}
\label{eq:U1qmCFT}
    i [H^{\CFT},q^{\CFT}(x)]=\partial_x m^{\CFT}(x).
\end{equation}

Similar to Eq.~\eqref{eq:LnCFT}, which defined the Virasoro generators, we may define the generators of the Kac--Moody algebra by
\begin{eqnarray}
\label{eq:U1JnCFT}
J^{\CFT}_n &=& \frac{1}{2\pi}\int_0^L dx\, e^{+inx\frac{2\pi}{L}} J^{\CFT}(x) \\
\bar{J}^{\CFT}_n &=& \frac{1}{2\pi}\int_0^L dx\, e^{-inx\frac{2\pi}{L}} \bar{J}^{\CFT}(x).\label{eq:U1JnCFTb}
\end{eqnarray}
They satisfy the $\mathfrak{u}(1)_k$ Kac--Moody algebra for some value $k\in \mathbb{Z}$, known as the level constant,
\begin{equation}
\label{eq:JmJn}
\begin{aligned}
&[J^{\CFT}_m, J^{\CFT}_n] =  km \delta_{m+n,0},\\
&[\bar{J}^{\CFT}_m, \bar{J}^{\CFT}_n] = km \delta_{m+n,0},\\
&[J^{\CFT}_m, \bar{J}^{\CFT}_n] = 0.
\end{aligned}
\end{equation}
The Virasoro generators and Kac--Moody generators satisfy the commutation relations
\begin{equation}
\label{eq:LmJn}
\begin{aligned}
[L^{\CFT}_m, J^{\CFT}_n] = -n J^{\CFT}_{m+n}, \\
[\bar{L}^{\CFT}_m, \bar{J}^{\CFT}_n] = -n \bar{J}^{\CFT}_{m+n},
\end{aligned}
\end{equation}
which is compatible with the fact that $J^{\CFT}$ is a Virasoro primary operator with conformal dimensions $(1,0)$. Setting $m=0$ in Eq.~\eqref{eq:LmJn}, we see that $J^{\CFT}_m$ changes the holomorphic dimension by $-m$, as what $L^{\CFT}_m$ does. Therefore, $J^{\CFT}_{m}$ is a raising operator with negative $m$, and a lowering operator with positive $m$. When $m=0$, $J^{\CFT}_{0}$ commutes with $L^{\CFT}_0$ and they have the same eigenstates. This reflects the fact that $J^{\CFT}(x)$ is a conserved current.
 
In analogy with the definition of Virasoro primary states in Eq.~\eqref{eq:primarydef}, we define Kac--Moody primary states,
\begin{equation}
       J^{\CFT}_n|\phi^{\CFT}_\alpha\rangle=0, ~~ \bar{J}^{\CFT}_n|\phi^{\CFT}_\alpha\rangle=0, ~~ \forall n>0.
\end{equation}
A Kac--Moody tower consists of a Kac--Moody primary state $|\phi^{\CFT}_\alpha\rangle$ with scaling dimension $\Delta_\alpha$ and conformal spin $s_\alpha$ and their descendent states, obtained from the primary state by sequentially acting with the generators. 

For instance, examples of Kac--Moody descendants include
\begin{equation}
|\phi^{\CFT}_\beta\rangle = J^{\CFT}_{-n} |\phi^{\CFT}_\alpha\rangle, ~~
|\phi^{\CFT}_\beta\rangle = \bar{J}^{\CFT}_{-n} |\phi^{\CFT}_\alpha\rangle , ~~ \text{for} \ n > 0. 
\end{equation}
with
\begin{equation} \label{eq:ladder}
\Delta_{\beta} = \Delta_{\alpha} + n , \quad s_{\beta} = s_{\alpha} \pm n, \quad \text{for} \ n > 0.
\end{equation}

It can be shown that all Kac--Moody primary states are also Virasoro primary states. However, there are Virasoro primary states that are not Kac--Moody primary states. For example,
\begin{eqnarray}
|J^{\CFT}\rangle &=& J^{\CFT}_{-1}|\mathbf{1}^{\CFT}\rangle \\
|\bar{J}^{\CFT}\rangle &=& \bar{J}^{\CFT}_{-1}|\mathbf{1}^{\CFT}\rangle \\
|J\bar{J}^{\CFT}\rangle &=& J^{\CFT}_{-1}\bar{J}^{\CFT}_{-1}|\mathbf{1}^{\CFT}\rangle,
\end{eqnarray}
that is, the states corresponding to the holomorphic current $J^{\CFT}$, the anti-holomophic current $\bar{J}^{\CFT}$ and their composite operator $J\bar{J}^{\CFT}$, which can be seen to be Virasoro primary states, all belong to the same Kac--Moody tower.

\subsection{Free compactified boson CFT}
A simple example of CFT that has Kac--Moody symmetry is a free compactified boson CFT, with action
\begin{equation}
    S^{\CFT}=\frac{1}{2}\int d\tau dx \, \left[ (\partial_\tau \phi)^2+(\partial_x \phi)^2 \right],
\end{equation}
where $\phi$ is identified with $\phi+2\pi R$, and $R$ is the \textit{compactification radius}.
The holomorphic and anti-holomorphic conserved currents read
\begin{equation}
    J^{\CFT}=i\partial\phi,~~ \bar{J}^{\CFT}=i\bar{\partial}\phi,
\end{equation}
where $\partial=(\partial_\tau+ i\partial_x)/2$ and $\bar{\partial}=(\partial_\tau- i\partial_x)/2$ are holomorphic and anti-holomorphic derivatives. In this CFT, the conservation relations $\bar{\partial}J^{\CFT}=0$ and $\partial \bar{J}^{\CFT}=0$ follow from the equation of motion $\partial\bar{\partial}\phi=0$. Now, using the definition of Kac--Moody generators in Eqs.~\eqref{eq:U1JnCFT}-\eqref{eq:U1JnCFTb} and the canonical commutation $[\phi(x),\partial_\tau\phi(y)]=\delta(x-y)$, we can verify the Kac--Moody algebra,
\begin{equation}\label{eq:u1comm}
\begin{aligned}
&[J^{\CFT}_m, J^{\CFT}_n] = [\bar{J}^{\CFT}_m, \bar{J}^{\CFT}_n] = m \delta_{m+n,0},\\
&[J^{\CFT}_m, \bar{J}^{\CFT}_n] = 0,
\end{aligned}
\end{equation}
which corresponds level constant $k=1$.

The free compactified boson CFT has infinitely many primary states with respect to the $\mathfrak{u}(1)_1$ Kac--Moody algebra. They are vertex operators $V^{\CFT}_{Q,M}$ labelled by integers $Q$ and $M$, with scaling dimensions and conformal spins
\begin{equation}
\Delta^{\CFT}_{V_{Q,M}} = \frac{Q^2}{R^2} + \frac{R^2M^2}{4}, \quad S^{\CFT}_{V_{Q,M}} = QM, \quad Q, M \in \mathbb{Z},
\end{equation}
where $Q$ and $M$ are the eigenvalues of $Q^{\CFT}$ and $M^{\CFT}$, respectively.

\begin{figure}
\begin{minipage}{0.78\linewidth}
\centering
\includegraphics[width=\linewidth]{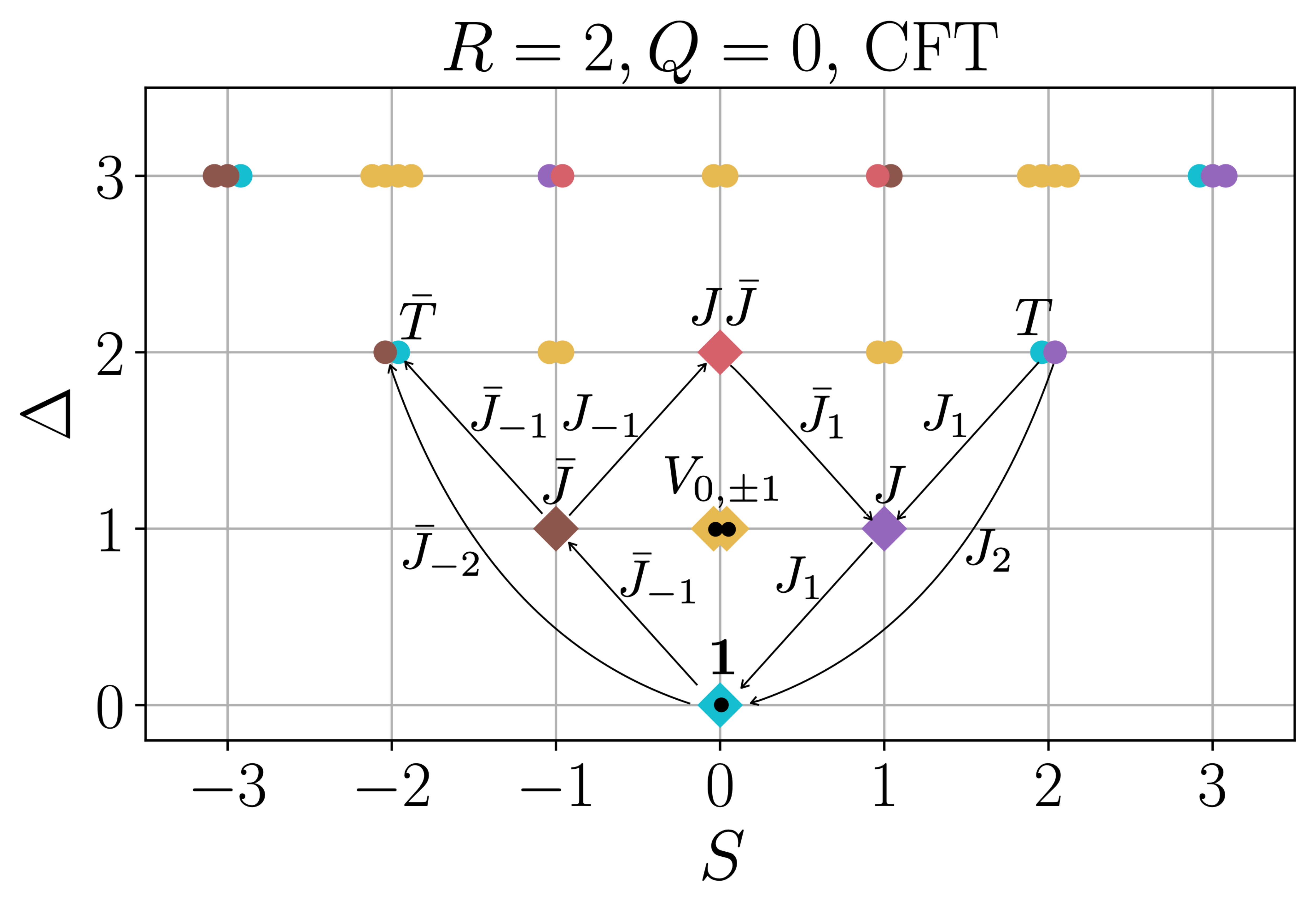}
\end{minipage}
\begin{minipage}{0.78\linewidth}
\centering
\includegraphics[width=\linewidth]{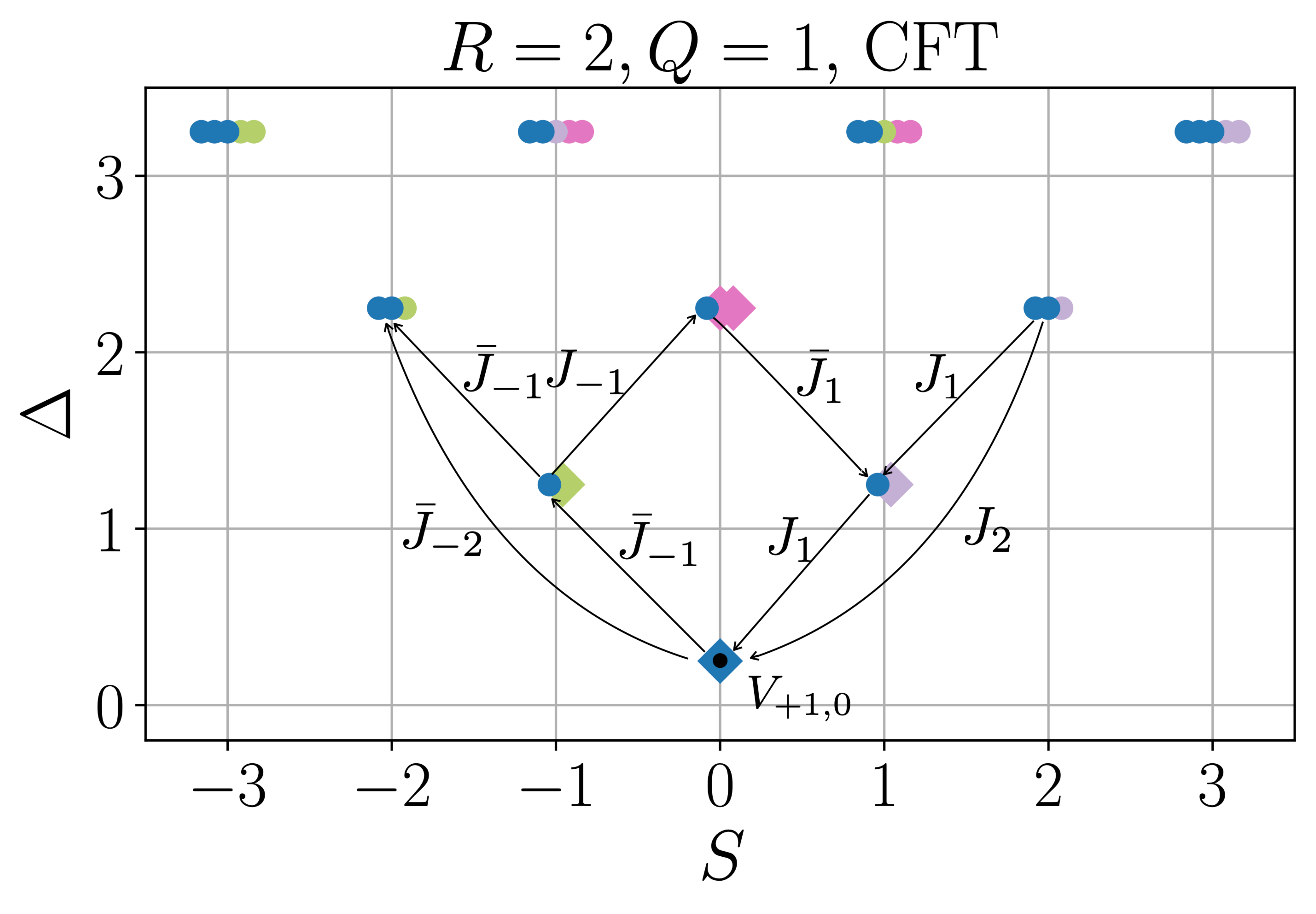}
\end{minipage}
\caption{Spectra of $Q=0$ sector (top) and $Q=1$ sector (bottom) of the free boson theory at compactification radius $R= 2$ and some examples of the actions of Kac--Moody generators $J_{n}^{\CFT}$ and $\bar{J}_{n}^{\CFT}$ on the low-energy eigenstates. States marked with diamonds are Virasoro primary states. Different Virasoro towers are shown in different colors. States marked with dots on top of the diamonds are the primary states with respect to the Kac--Moody symmetry in these two sectors. For the purpose of visibility, we shifted states horizontally to better show the energy-momentum degeneracies. It should be kept in mind that in these figures the conformal spin $S$ is quantized to only take exact integer numbers. }
\label{fig:boson}
\end{figure}

\subsection{Lattice Kac--Moody generators}

Consider now a critical quantum spin chain on the circle with Hamiltonian $H$ and global $U(1)$ symmetry (see e.g. the XXZ model below). Our goal is to construct lattice operators $J_n,\bar{J}_m$ that correspond to the Kac--Moody generators $J^{\CFT}_n,\bar{J}^{\CFT}_m$. On the lattice, let the conserved $U(1)$ charge be
\begin{equation}
    Q=\sum_{j} q_j,
\end{equation}
which commutes with the Hamiltonian $[H, Q] =0$. We identify $Q$ with the CFT charge operator $Q^{\CFT}$ in Eq.~\eqref{eq:U1QCFT}, and $q$ with the CFT current density $q^{\CFT}$ in Eq.~\eqref{eq:U1qCFT}. 
In order to find the Kac--Moody generators, one also needs to find the lattice operator $m$ corresponding to $m^{\CFT}$ in Eq.~\eqref{eq:U1mCFT}. 

Our proposal is that the corresponding lattice operators should satisfy an equation analogous to Eq.~\eqref{eq:U1qmCFT}, after replacing the spatial derivative $\partial_x$ with a finite difference, see Eq.~\eqref{eq:U1qm} below. Then by Eqs.~\eqref{eq:U1qCFT}-\eqref{eq:U1mCFT}, the lattice operators $J_j$ and $\bar{J}_j$ that correspond to $J^{\CFT}$ and $\bar{J}^{\CFT} $ can be identified as in Eq.~\eqref{eq:U1latticeJi} below. Finally the lattice operators $J_n$ and $\bar{J}_n$ corresponding to the Kac--Moody generators are constructed as Fourier modes in Eq.~\eqref{eq:U1J}. 

\begin{empheq}[box=\fbox]{align}
\label{eq:U1qm}
&i[H, q_j] = m_{j} - m_{j-1} \\
\label{eq:U1latticeJi}
&J_j = \frac{q_j + m_j}{2} , \quad \bar{J}_j = \frac{q_j - m_j}{2} \\
\label{eq:U1J}
&J_n = \sum_{j}^{N} e^{i j n \frac{2\pi}{N}} J_{j}, \quad 
\bar{J}_n = \sum_{j}^{N} e^{- i j n \frac{2\pi}{N}}  \bar{J}_{j}. 
\end{empheq}

Eqs.\eqref{eq:U1qm}, \eqref{eq:U1latticeJi} and \eqref{eq:U1J} are the main proposal of this paper (together with their generalization below to the non-Abelian case). 

Let us elaborate on the form of $m_j$ when $H$ is a nearest-neighbor and a next-to-nearest-neighbor Hamiltonian. These specializations will be useful in the applications to specific lattice models.
\subsubsection{Nearest-neighbor Hamiltonian}
Consider a nearest-neighbor Hamiltonian $H = \sum_{j} h_{j,j+1}$ with a global symmetry that is realized on-site with charge $Q = \sum_{j} q_j$. If follows from $[H, Q]=0$ that at for any pair $(j, j+1)$ of contiguous sites, 
\begin{equation}
[h_{j, j+1}, q_j + q_{j+1}] = 0.
\end{equation}
The LHS of Eq.~\eqref{eq:U1qm} then becomes
\begin{equation}
\begin{aligned}
i [H, q_j] &= i ([h_{j-1,j}, q_j] + [h_{j, j+1}, q_j]) \\
 &= i (-[h_{j-1,j}, q_{j-1}] + [h_{j, j+1}, q_j]),
\end{aligned}
\end{equation}
and matching it with the RHS of Eq.~\eqref{eq:U1qm}, we obtain
\begin{equation}\label{eq:2local}
m_j = i[h_{j,j+1}, q_j],
\end{equation}
which is our proposal as a lattice version of the locally conserved current $m^{\CFT}(x)$.

\subsubsection{Next-to-nearest-neighbor Hamiltonian}
As a second example, consider a next-to-nearest-neighbor Hamiltonian $H = \sum_j h_{j-1, j, j+1}$, again with a global symmetry that is realized on-site with charge $Q = \sum_{j} q_j$. To simplify the notation, in the following $h_{(j)} \coloneqq h_{j-1,j,j+1}$.

At any set $(j-1,j,j+1)$ of three consecutive sites,
\begin{equation}
[h_{(j)}, q_{j-1} + q_j + q_{j+1}] = 0,
\end{equation}
we can check that
\begin{equation}
\begin{aligned}
i\left[H, \frac{q_{j}+q_{j+1}}{2}\right] 
= \frac{1}{2}(m_{j+1} - m_{j-1})
\end{aligned}
\end{equation}
where 
\begin{equation}\label{eq:3local}
m_{j} = i([h_{(j+1)}, q_j] - [h_{(j)}, q_{j+1}]).
\end{equation}
This is a discrete version of Eq.~\eqref{eq:U1qmCFT} in the case of next-to-nearest-neighbor Hamiltonian. Therefore we identify $m_j$ in Eq.~\eqref{eq:3local} with $m^{\CFT}(x)$.

To reiterate, given a local Hamiltonian $H$ with a microscopic global symmetry realized on-site, that is with charge $Q=\sum_j q_j$, our proposal gives a concrete way of constructing the lattice current density $m_j$ via Eq.~\eqref{eq:U1qm} and subsequently lattice Kac--Moody generators $J_n$ and $\bar{J}_n$ via Eq.~\eqref{eq:U1latticeJi}-\eqref{eq:U1J}. 


\subsection{Example: XXZ model}
As a test of our construction of lattice current generators $J_n$ and $\bar{J}_{n}$ for a critical quantum spin chain with a global $U(1)$ symmetry, we consider the XXZ model 
\begin{equation} \label {eq: XXZ}
H = - \frac{2\gamma}{\pi \sin{\gamma}} \sum_{j=1}^{N}\left(
S_{j}^{X}S_{j+1}^{X} + S_{j}^{Y}S_{j+1}^{Y}  - \cos{\gamma}S_{j}^{Z}S_{j+1}^{Z}
\right)
\end{equation}
with anisotropy $\gamma \in [0, \pi )$ and $S^\alpha = \sigma^\alpha/2$, where $\sigma^{x}, \sigma^y$ and $\sigma^z$ stand for the Pauli matrices. 
This model displays indeed a global $U(1)$ symmetry realized on-site, generated by the charge operator $Q = \sum_{j=1}^{N} S_{j}^{Z}$, which commutes with the Hamiltonian, $[Q, H] = 0$. In the continuum limit the XXZ model is equivalent via bosonization to the Gaussian model of a free massless boson compactified on circle with radius $R = \sqrt{2\pi/(\pi - \gamma)}$ and the symmetry algebra in the corresponding CFT theory is the Kac--Moody algebra with $\mathfrak{g} = \mathfrak{u}(1)$. 

As discussed in the last section, we use Eq.~\eqref{eq:2local} to compute the lattice current $m_j$, 
\begin{equation}\label{eq:u1m}
m_j =  \frac{2 \gamma}{\pi \sin{\gamma}} \left(
S_{j}^{X}S_{j+1}^{Y} - S_{j}^{Y}S_{j+1}^{X}
\right) .
\end{equation}
We note that the lattice currents $q_j$ and $m_j$ for the XXZ model can also be understood using bosonization, as discussed in Appendix~\ref{appendix:a}. 

Following Eq.~\eqref{eq:J} the lattice versions of Kac--Moody generators can be constructed as

\begin{equation}
\label{eq:Jnxxz}
\begin{aligned}
J_n =  \sum_{j=1}^{N} e^{+ijn \frac{2\pi}{N}}  \frac{1}{2}
\left[
S_{j}^{Z}+  \frac{2\gamma}{\pi \sin{\gamma}}  \left(
S_{j}^{X}S_{j+1}^{Y} - S_{j}^{Y}S_{j+1}^{X}  
\right)
\right], \\
\bar{J}_n =  \sum_{j=1}^{N} e^{-ijn \frac{2\pi}{N}}  \frac{1}{2}
\left[
{S_{j}^{Z}} -  \frac{2\gamma}{\pi \sin{\gamma}}  \left(
S_{j}^{X}S_{j+1}^{Y} - S_{j}^{Y}S_{j+1}^{X}  
\right)
\right] .
\end{aligned}
\end{equation}

\begin{figure}[h]
\begin{minipage}{0.78\linewidth}
\centering
\includegraphics[width=\linewidth]{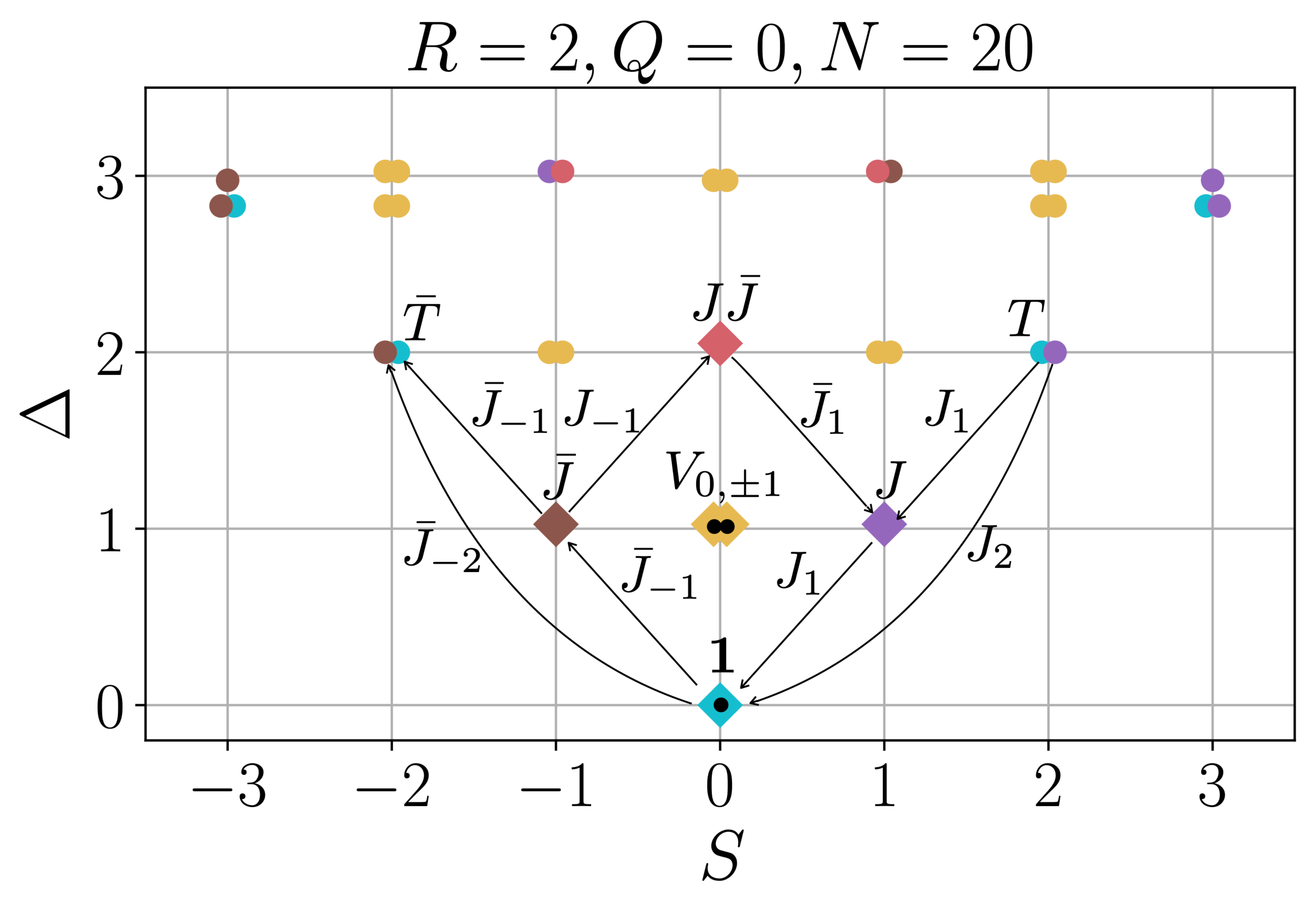}
\end{minipage}
\begin{minipage}{0.78\linewidth}
\centering
\includegraphics[width=\linewidth]{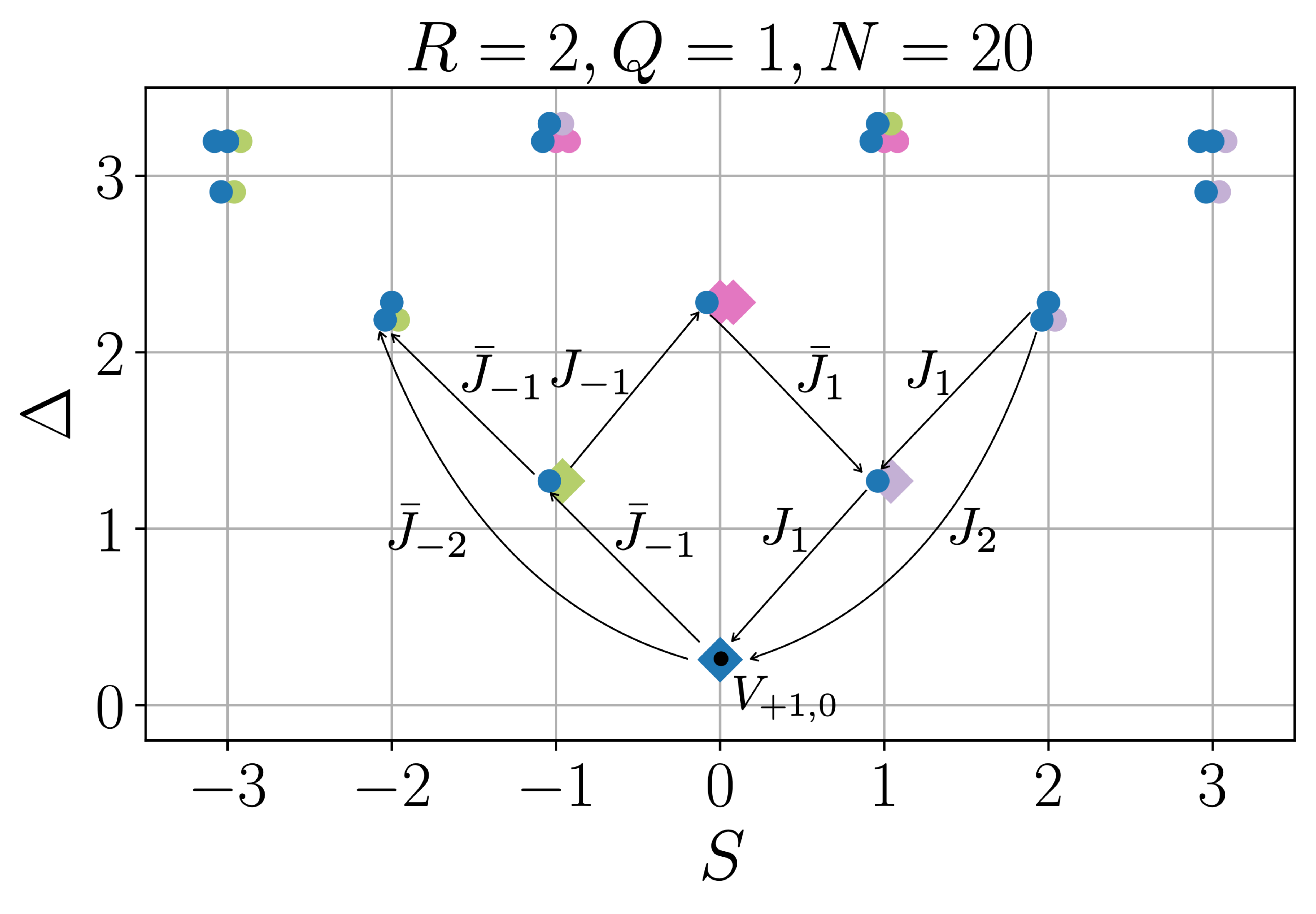}
\end{minipage}
\caption{ Spectra of $Q=0$ sector and $Q=1$ sector of the XXZ model with $\gamma = \pi/2 $ at system size $N = 20$. We find the spectrum here in good agreement with the exact free boson spectrum in Fig.~\ref{fig:boson}. Kac--Moody primary states are labeled with dots. Some examples of actions of $J_n$ and $\bar{J}_n$ are illustrated.  }
\label{fig:spectrum}
\end{figure}

To numerically check our proposal, we use exact diagonalization to find a set of low energy eigenstates $\ket{\psi_\alpha}$. We simultaneously diagonalize the Hamiltonian and the translation operator with periodic boundary conditions to get the scaling dimensions $\Delta_{\alpha}$ and conformal spins $s_{\alpha}$. We also sort the low-energy eigenstates into different sectors according to the eigenvalues of $Q=\sum_j S^Z_j$. As an example, in Fig. \ref{fig:spectrum} the low-energy spectra of XXZ spin chain of 20 sites with $\gamma = \pi/2$ in the $Q = 0$ sector and $Q=1$ sector are plotted.  

\begin{figure}[h]
\centering
\includegraphics[width=\linewidth]{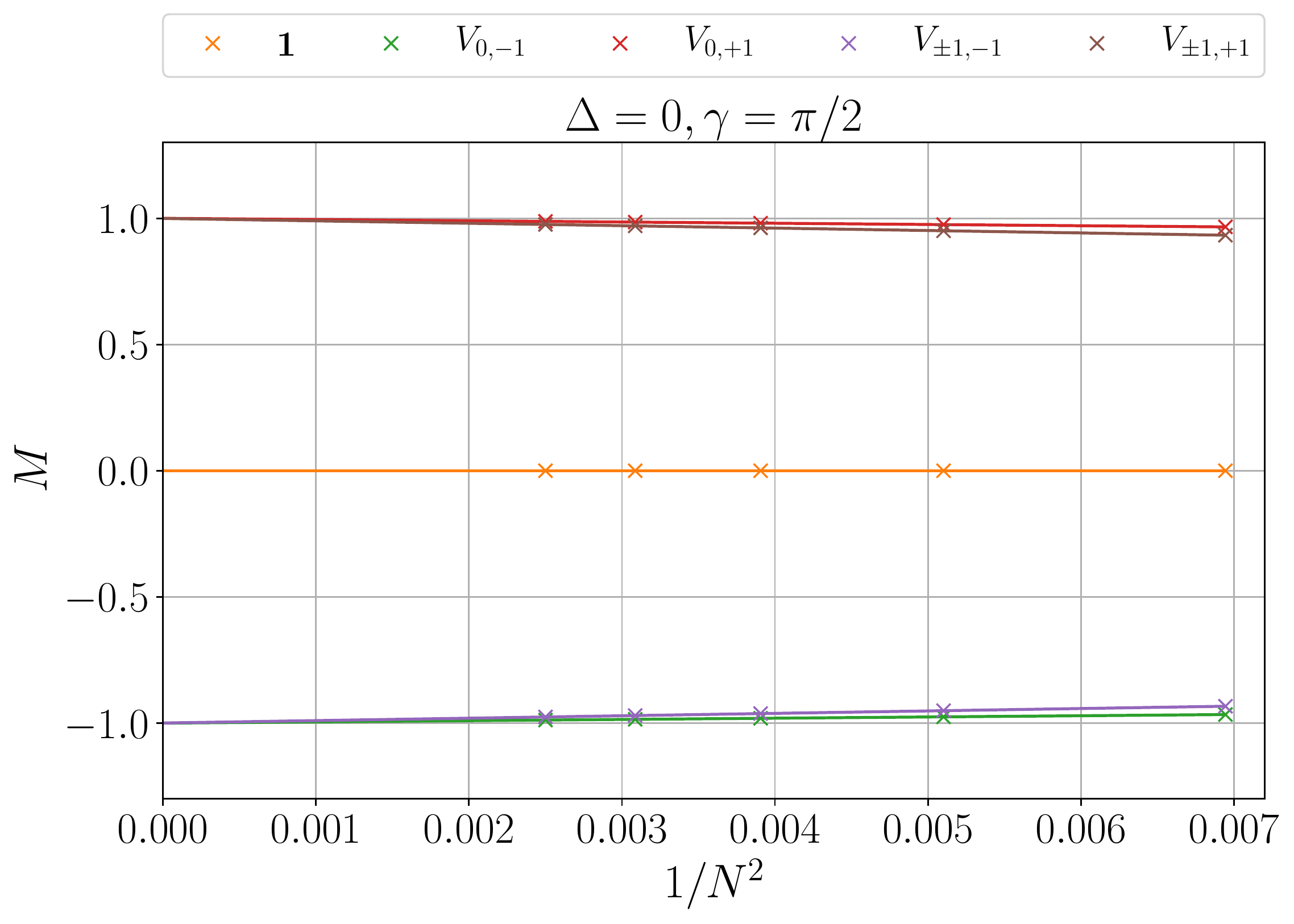}
\caption{Expectation values of $M$ in the eigenbasis of $Q$ approach integers in the thermodynamic limit. Finite size scaling of $\mel{\psi}{M}{\psi} $ with $\ket{\psi} = \ket{\mathbf{1}}, \ket{V_{0,{\pm1}}}, \ket{V_{\pm1, \pm2}}$ at system sizes $N=12, 14, 16, 18, 20$. Results for different compatification radii can be found in Appendix~\ref{appendix:c}.}
\label{fig:xx_m}
\end{figure}

That the above lattice generators $J_n$ and $\bar{J}_n$ approximately satisfy the Kac--Moody algebra at low energies can be verified by directly evaluating Eqs.~\eqref{eq:JmJn}-\eqref{eq:LmJn} on low energy states. Those constitute an infinite set of conditions. In the following, we will demonstrate it for a subset of these conditions that are physically important. Other conditions can be confirmed similarly.
From Eq.~\eqref{eq:LmJn} for $m=n=0$, we see that the charge $M = J_0 - \bar{J}_0$ should approximately satisfy $[M,H]=0$ and $[M,Q]=0$ when acting on low-energy states. While for general anisotropy $\gamma$ the operator $M=\sum_j m_j$ does not necessarily commute with the Hamiltonian, in the specific case of the XX model ($\gamma = \pi/2$), $M \sim \sum_j S_j^X S_{j+1}^Y - S_j^Y S_{j+1}^X $ does commute with $H$. In fact, there exists a series of conserved charges in the XX model \cite{GRABOWSKI1995299}. However $M$ does not exactly commute with $Q$ on the lattice. To confirm that the microscopic global $U(1)$ symmetry turns indeed into an emergent $U(1)\times U(1)$ symmetry at low energies, we check both the commutation relation $\mel{\psi}{[Q,M]}{\psi}$ and the expectation value $\mel{\psi}{M}{\psi}$ in the eigenbasis of $Q$. For low energy eigenstates, we find that $Q$ and $M$ commute, up to a finite size error. We further show that the expectation value of the charge $M$ approaches an integer value, see Fig.~\ref{fig:xx_m}. This is compatible with the emergence of a global $U(1)\times U(1)$ symmetry and allows us to identify vertex operators $V_{Q, M}$ according to their eigenvalues of $Q$ and $M$. 

\begin{figure}
\center \includegraphics[width=\linewidth]{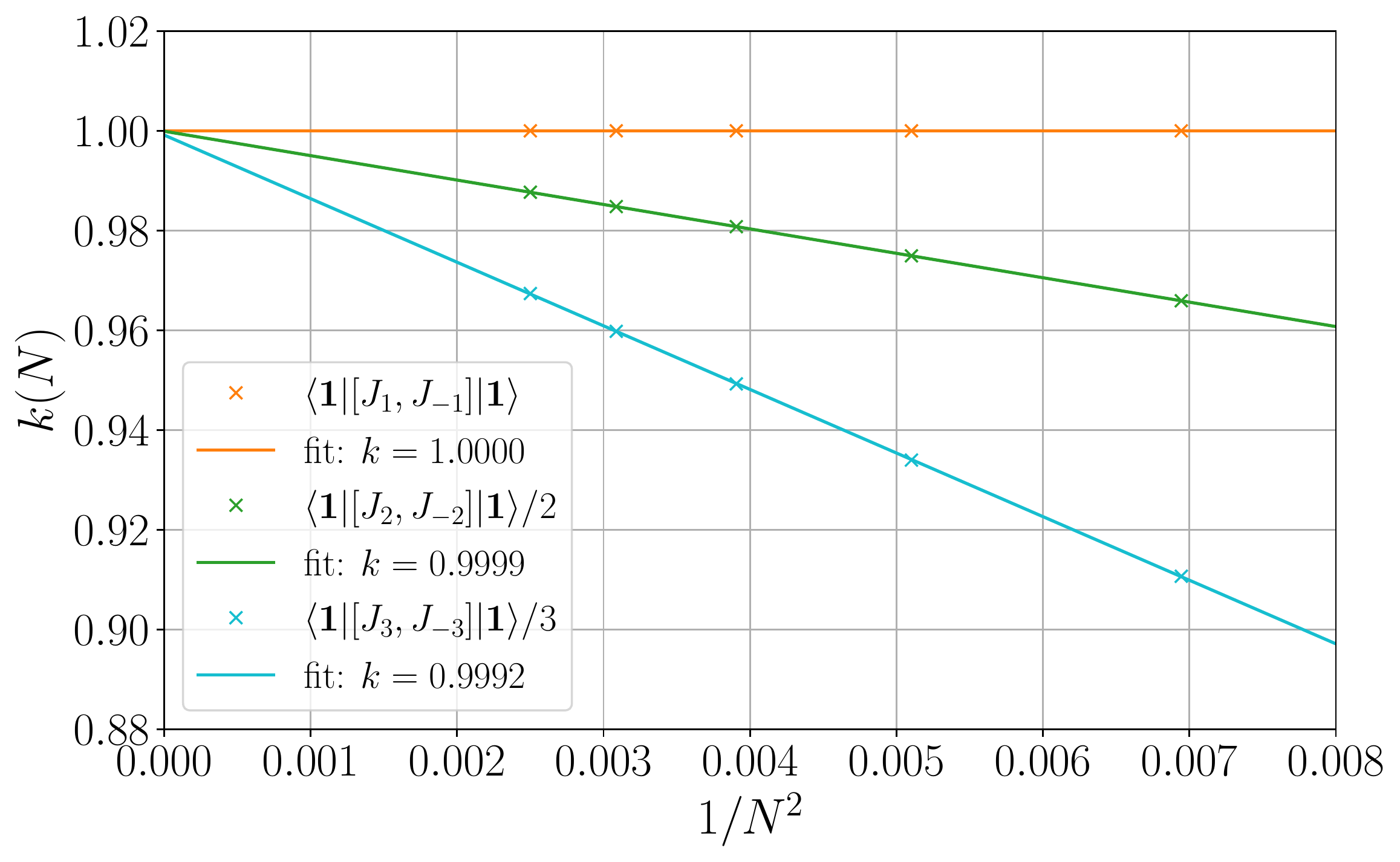}
\caption{Extracting level constant from matrix elements $k_1 = \mel{\mathbf{1}}{[J_1, J_{-1}]}{\mathbf{1}}$ ,  $k_2 = \frac{1}{2} \mel{\mathbf{1}}{[J_2, J_{-2}]}{\mathbf{1}}$, and  $k_3 = \frac{1}{3} \mel{\mathbf{1}}{[J_1, J_{-1}]}{\mathbf{1}}$ at system sizes $N = 12, 14, 16, 18, 20$.}
\label{fig:kfit}
\end{figure}

We can also examine the commutation relations in Eq.~\eqref{eq:JmJn} in the energy-momentum eigenbasis by calculating matrix elements $\mel{\psi}{[J_n, J_m]}{\psi} $. The level constant $k$ can be extracted from $\mel{\psi }{[J_n, J_{-n}]}{\psi}$, according to Eq.~\eqref{eq:u1comm}. For example, we can estimate $k$ using the three expectation values $k_1 = \mel{\mathbf{1}}{[J_1, J_{-1}]}{\mathbf{1}}$, $k_2 = \frac{1}{2} \mel{\mathbf{1}}{[J_2, J_{-2}]}{\mathbf{1}}$ and  $k_3 = \frac{1}{3} \mel{\mathbf{1}}{[J_3, J_{-3}]}{\mathbf{1}}$, as shown in Fig.~\ref{fig:kfit}. Linear extrapolation gives $ k_1 = 1.0000$, $k_2 = 0.9999$ and $k_3 = 0.9992$, which are in good agreement with theoretical value $k=1$.

We check the action of $J_n$ on low energy eigenstates by examining matrix elements $\mel{\psi_\alpha}{ J_n }{ \psi_\beta} $ (similarly for the anti-chiral part $\bar{J}_n$). By numerically checking whether a state is annihilated by $J_{1}, \bar{J}_{1}$ and $J_{2}, \bar{J}_{2}$, up to finite size error, we successfully identify all the Kac--Moody primaries. In Fig. \ref{fig:spectrum}, the candidate primary states in the $Q=0$ and $Q=1$ sectors are marked.  Compared with the analytical result of free boson CFT in Fig. \ref{fig:boson}, our lattice generators identify the primary states correctly. By acting on a primary state with the lattice generators $J_{n}$ and $\bar{J}_{n}$, we can obtain all the descendent states of that primary state. Up to finite size error, $J_n$ and $\bar{J}_n$ act as ladder operators - they change the scaling dimension and conformal spin as illustrated in the example of Eq.~\eqref{eq:ladder}. We plot a few examples of their actions in Fig. \ref{fig:spectrum}.  Together we are able to organize all low-energy eigenstates into different Kac--Moody towers. 

In addition to the case of XX model with $\gamma = \pi/2$, our proposal can be verified for general XXZ model. We provide numerical results for some other values of $\gamma$ in Appendix \ref{appendix:c}. Up to an overall normalization constant of $J$ and $\bar{J}$, we find consistently $k=1$ and expectation values of $M$ to be integers, confirming the Kac--Moody algebra.

\section{Lattice realization of Kac--Moody algebra: the $SU(2)$ case}
\label{sec:su2}
In this section we generalize the construction of lattice Kac--Moody generators to general Lie groups. We will first introduce the Kac--Moody algebra with general semisimple Lie groups. Then we consider a specific example, the free compactified boson with compactification radius $R=\sqrt{2}$, which possesses an $\mathfrak{su}(2)_1$ Kac--Moody symmetry. We then consider critical quantum spin chains with semisimple Lie group symmetry and construct approximate lattice Kac--Moody generators. Finally we present the numerical verification with the Heisenberg model with next-to-nearest neighbor interactions.

\subsection{Kac--Moody algebra with a non-Abelian group}

We consider a CFT that has an internal semisimple Lie group symmetry $G$, where the Lie algebra is denoted as $\mathfrak{g}$. The extended symmetry is generated by \textit{Kac--Moody generators} that form the \textit{Kac--Moody algebra}, denoted as $\hat{\mathfrak{g}}_k $, where $k$ is the \textit{level constant} of the Kac--Moody algebra. In a CFT with Kac--Moody algebra, there exists a set of chiral and anti-chiral current operators $J^{\alpha,\CFT}$ and $\bar{J}^{\alpha,\CFT}$ with conformal dimensions $(1,0)$ and $(0,1)$, where $\alpha=1,2,\cdots,\mathrm{dim}~\mathfrak{g}$ labels different currents.

The Kac--Moody generators are Fourier modes of the currents,
\begin{eqnarray}
\label{eq:JnCFT}
J^{\alpha,\CFT}_n &=& \frac{1}{2\pi}\int_0^L dx\, e^{+inx\frac{2\pi}{L}} J^{\alpha,\CFT}(x) \\
\bar{J}^{\alpha,\CFT}_n &=& \frac{1}{2\pi}\int_0^L dx\, e^{-inx\frac{2\pi}{L}} \bar{J}^{\alpha,\CFT}(x).
\end{eqnarray}
They satisfy the Kac--Moody algebra $\hat{\mathfrak{g}}_k $,

\begin{equation} \label{eq:comrel}
\begin{aligned}
&[\J{m}{\alpha},\J{n}{\beta}] = i \sum_\gamma f^{\alpha \beta \gamma} \J{m+n}{\gamma} + k m \delta^{\alpha \beta} \delta_{m+n, 0}\\
&[\J{m}{\alpha} , \Jbar{n}{\beta}] = 0\\
&[\Jbar{m}{\alpha}, \Jbar{n}{\beta}] = i \sum_\gamma f^{\alpha \beta \gamma} \Jbar{m+n}{\gamma} + k m \delta^{\alpha \beta} \delta_{m+n, 0}
\end{aligned}
\end{equation}
where $f^{\alpha \beta \gamma}$ are the structure constants of the Lie algebra $\mathfrak{g}$ and $k$ is the level constant. 

Setting $n=m=0$ in Eq.~\eqref{eq:comrel}, we see that the zero modes of the currents form exactly the ordinary Lie algebra $\mathfrak{g}$ of the global symmetry $G$,
\begin{equation}
[\J{0}{\alpha} , \J{0}{\beta}] 
= i \sum_\gamma f^{\alpha \beta \gamma} \J{0}{\gamma}.
\label{eq:j0comrel}
\end{equation}
Since $J^{\alpha,\CFT}_0$ and $\bar{J}^{\alpha,\CFT}_0$ are separately conserved, the global symmetry is $G\times G$, as it also happened in the Abelian $U(1)$ case of the previous section.

The holomorphic and anti-holomorphic currents transform into each other under spatial parity. We may consider the linear combinations of currents that have definite parity, 
\begin{eqnarray}
\label{eq:qCFT}
q^{\alpha, \CFT}(x) &\equiv& \J{}{\alpha}(x) + \Jbar{}{\alpha}(x) \\
\label{eq:mCFT}
m^{\alpha, \CFT}(x) &\equiv& \J{}{\alpha}(x) - \Jbar{}{\alpha}(x).
\end{eqnarray}
They correspond to charges,
\begin{eqnarray}
Q^{\alpha, \CFT} &\equiv& \J{0}{\alpha} + \Jbar{0}{\alpha} \\
M^{\alpha, \CFT} &\equiv& \J{0}{\alpha} - \Jbar{0}{\alpha}.
\end{eqnarray}
From Eq.~\eqref{eq:j0comrel}, it follows that charges $Q^{\alpha, \CFT}$ satisfy the Lie algebra $\mathfrak{g}$
\begin{equation}
[Q^{\alpha, \CFT}, Q^{\beta, \CFT}] 
= i \sum_\gamma f^{\alpha \beta \gamma} Q^{\gamma, \CFT}.
\end{equation}  

\subsection{Free compactified boson at $R=\sqrt{2}$}
At $R=\sqrt{2}$, the free compactified boson has three Virasoro primary operators with conformal dimension $(1,0)$,
\begin{eqnarray}
J^{x,\CFT} &=& \frac{1}{\sqrt{2}}(V^{\CFT}_{1,1}+V^{\CFT}_{-1,-1})       \\
J^{y,\CFT} &=& \frac{1}{\sqrt{2}i}(V^{\CFT}_{1,1}-V^{\CFT}_{-1,-1})       \\
J^{z,\CFT} &=& i\partial\phi.
\end{eqnarray}

Together with three anti-holomorphic currents (with similar expressions), their Fourier modes constitute the $\mathfrak{su}(2)_1$ algebra,

\begin{equation}\label{eq:struct}
\begin{aligned}
&[J^{\alpha,\CFT}_m, J^{\beta,\CFT}_n] = \sum_{\gamma} i\sqrt{2}\epsilon^{\alpha\beta\gamma}J^{\gamma,\CFT}_{n+m}+m \delta_{m+n,0} \delta^{\alpha \beta},\\
&[J^{\alpha,\CFT}_m, \bar{J}^{\beta,\CFT}_n] = 0 \\
&[\bar{J}^{\alpha,\CFT}_m, \bar{J}^{\beta,\CFT}_n] =\sum_{\gamma} i\sqrt{2}\epsilon^{\alpha\beta\gamma}\bar{J}^{\gamma,\CFT}_{n+m}+m \delta_{m+n,0} \delta^{\alpha \beta} ,
\end{aligned}
\end{equation}
where we can read off the structure constants $f^{\alpha\beta\gamma}=\sqrt{2}\epsilon^{\alpha\beta\gamma}$ and the level constant $k=1$. For later convenience, we also define 
\begin{equation}
\begin{aligned}
        &J^{\pm,\CFT}_n = \frac{1}{\sqrt{2}}(J^{x,\CFT}_n\pm i J^{y,\CFT}_n),\\
        &J^{3, \CFT}_n = \frac{1}{\sqrt{2}}J^{z,\CFT}_n
\end{aligned}
\end{equation}

\begin{figure}
\begin{minipage}[b]{0.78\linewidth}
\centering
\includegraphics[width=\textwidth]{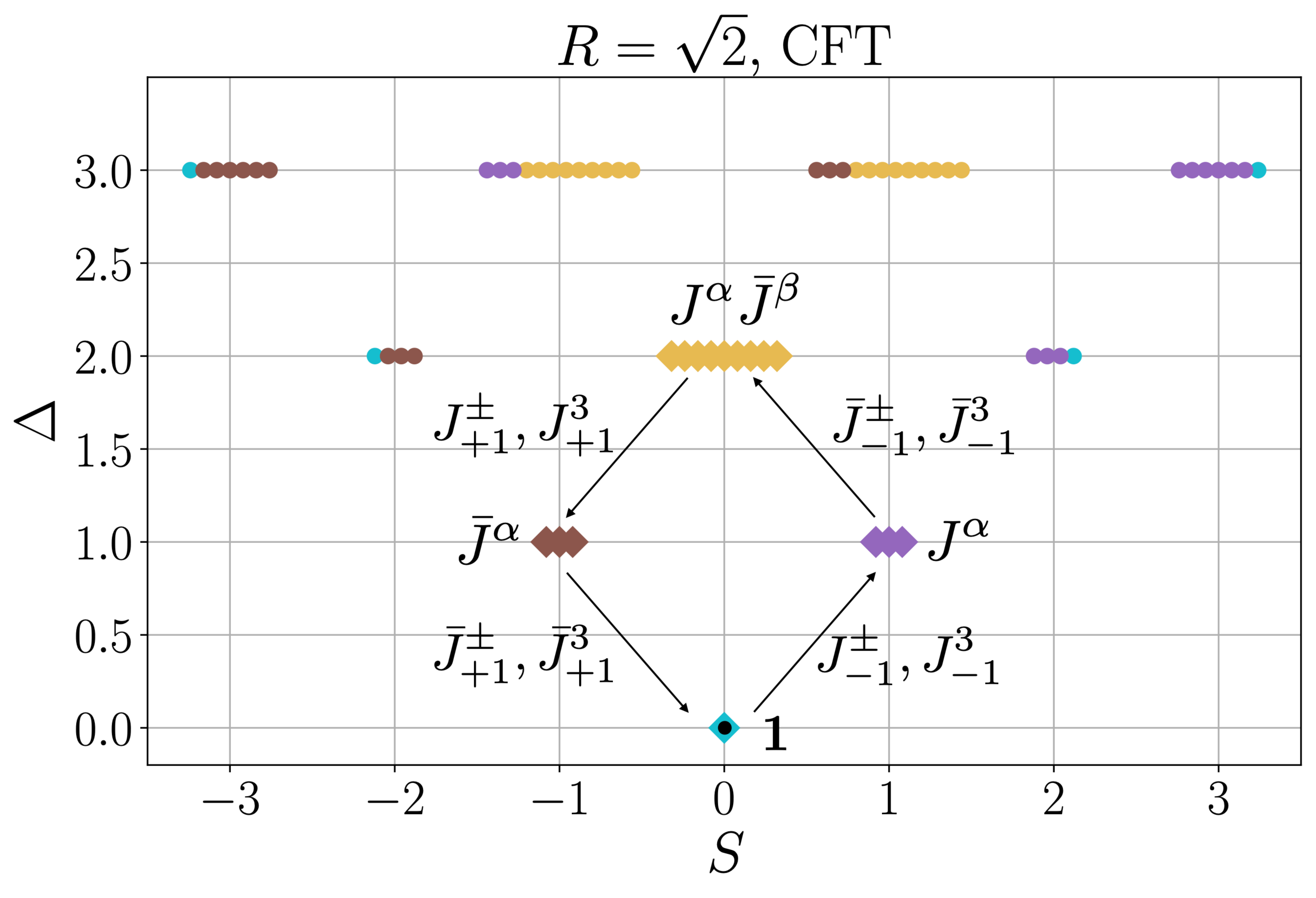}
\end{minipage}
\begin{minipage}[b]{0.78\linewidth}
\centering
\includegraphics[width=\textwidth]{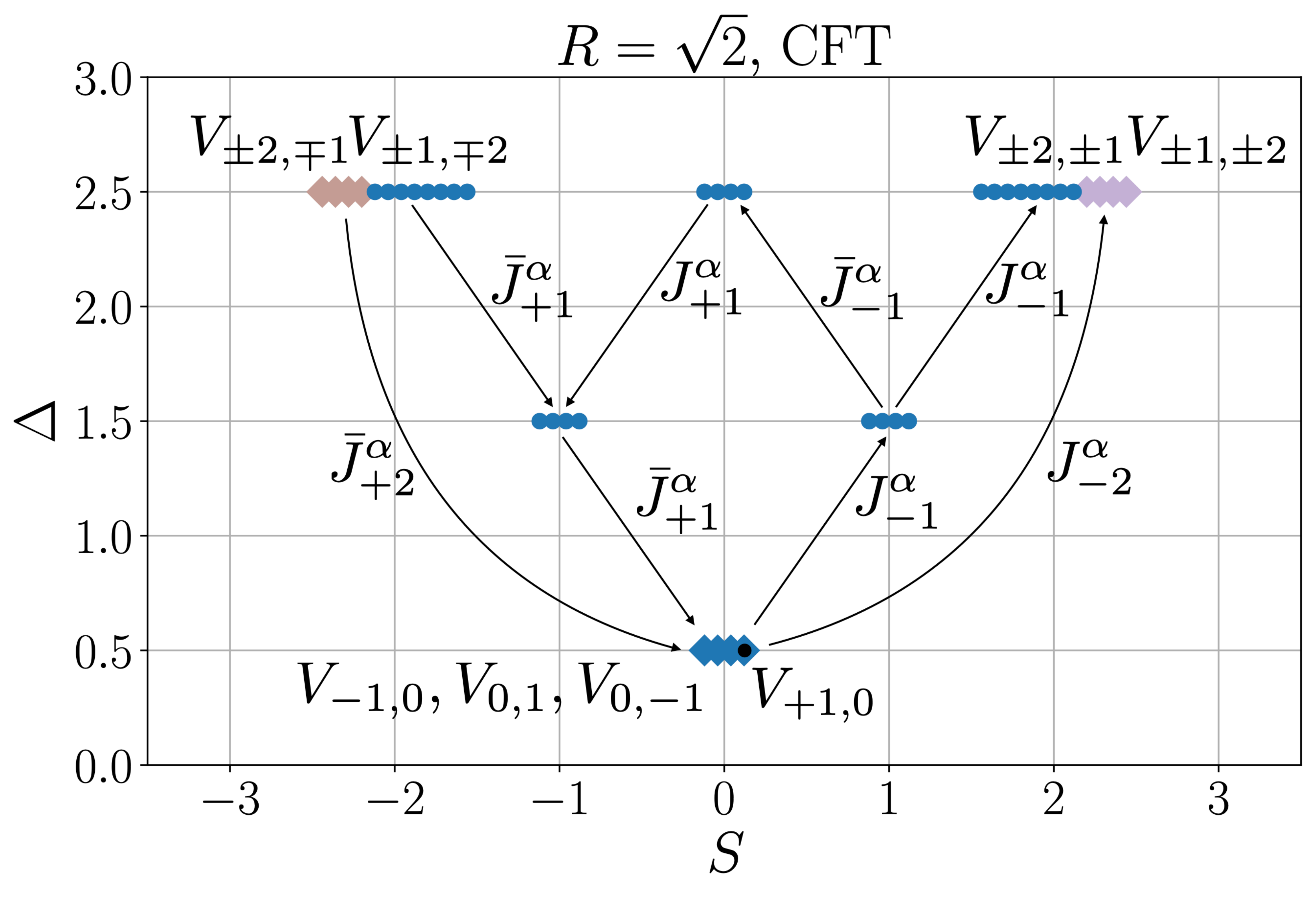}
\end{minipage}
    \caption{Exact spectrum of the free boson at compatification radius $R = \sqrt{2}$ divided into two Kac--Moody towers. Virasoro primary states are labeled with diamonds, among which Kac-Moody primary states are marked with dots. Top: Kac--Moody tower of the identity operator $|\mathbf{1}^{\CFT}\rangle$. The current operators $J^\alpha, \bar{J}^\beta, J^\alpha \bar{J}^\beta$ are labeled. Bottom: Kac--Moody tower of operator $|V^{\CFT}_{1,0}\rangle$. Some vertex states $V_{m,n}$ are labeled. Some examples of actions of $J_{n}^{\alpha, \CFT}$ are illustrated with arrows.}
    \label{fig:XXX_CFT01}
\end{figure}

While there can be infinite (but countable) Virasoro primary states, in the presence of additional symmetry, the number of primary states with respect to the extended symmetry can be finite. In the case of $\mathfrak{su}(2)_1$, a Kac--Moody primary state $\ket{j^\CFT}$ satisfies
\begin{equation}
\label{eq:hws}
    \begin{aligned}
        J_{n}^{\alpha, \CFT} \ket{j^\CFT} &= 0 \quad \text{for}\quad  n>0, \\
        J_0^{+, \CFT} \ket{j^\CFT} &= 0,\\
        J_0^{3, \CFT} \ket{j^\CFT} &= j\ket{j^\CFT}.
    \end{aligned}
\end{equation}
where $j = 0, \frac{1}{2}$.  
The only two $SU(2)$ level 1 Kac--Moody primary operators are the identity operator $\mathbf{1}^{\CFT}$ with $\Delta^{\CFT}=0$ and $s^{\CFT}=0$ and $V^{\CFT}_{1,0}$ with $\Delta^{\CFT}=1/2$ and $s^{\CFT}=0$. Other Virasoro primary operators can be obtained by acting with Kac--Moody generators on these primaries, as in the examples shown in Fig.~\ref{fig:XXX_CFT01}. For instance, in the Kac--Moody tower of the identity, $|V^{\CFT}_{1,1}\rangle$ and $|V^{\CFT}_{2,0}\rangle$ can be obtained in the following way,
\begin{eqnarray}
\label{eq:kmtower1}
 J^{+,\CFT}_{-1}|\mathbf{1}^{\CFT}\rangle=|V^{\CFT}_{1,1}\rangle  \\
 \label{eq:kmtower2}
 \bar{J}^{+,\CFT}_{-1}|V^{\CFT}_{1,1}\rangle=|V^{\CFT}_{2,0}\rangle.
\end{eqnarray}
Other Virasoro primary operators with scaling dimensions $1$ or $2$ can be obtained in an analogous way. In the Kac--Moody tower of $V_{1,0}$, 
\begin{eqnarray}
\label{eq:kmtower3}
  J^{+,\CFT}_{0}|V^{\CFT}_{-1,0}\rangle=|V^{\CFT}_{0,1}\rangle \\
  \label{eq:kmtower4}
 J^{-,\CFT}_{0}|V^{\CFT}_{1,0}\rangle=|V^{\CFT}_{0,-1}\rangle \\
 \label{eq:kmtower5}
 J^{+,\CFT}_{-2}|V^{\CFT}_{1,0}\rangle=|V^{\CFT}_{2,1}\rangle \\
 \label{eq:kmtower6}
 J^{-,\CFT}_{-2}|V^{\CFT}_{-1,0}\rangle=|V^{\CFT}_{-2,-1}\rangle.
\end{eqnarray}
Other Virasoro primary operators with scaling dimension $1/2$ or $5/2$ can be obtained in a similar way. These relations can be justified by the operator product expansion of vertex operators (see appendix \ref{appendix:b}). 

\subsection{Lattice Kac--Moody generators for general $G$}
 Consider a lattice model with Hamiltonian $H$ and an on-site symmetry $G$ with Lie algebra $\mathfrak{g}$. The symmetry is associated with a set of conserved charges
\begin{equation}
Q^{\alpha} =\sum_{i}^{N} q^{\alpha}_i, ~~\alpha=1,2,\cdots,\mathrm{dim}~\mathfrak{g}.
\end{equation}
The charges commute with the Hamiltonian, $[H, Q^{\alpha}] = 0$, and satisfy 
\begin{equation}
[Q^{\alpha} , Q^{\beta}] = i  f^{\alpha \beta \gamma} Q^{\gamma}.
\end{equation}
The one-site lattice operator $q^{\alpha}_j$ corresponds to the CFT operator $q^{\alpha,\CFT}(x)$ in Eq.~\eqref{eq:qCFT}, because they are both conserved and have even spatial parity. As in the $U(1)$ case, we need to find the lattice operator $m^{\alpha}_j$ corresponding to $m^{\alpha,\CFT}(x)$ in Eq.~\eqref{eq:mCFT}.

The derivation that leads to Eq.~\eqref{eq:U1qmCFT} goes through for each conserved current $J^{\alpha,\CFT}(x)$, and we obtain
\begin{equation}
\label{eq:qmCFT}
    i [H^{\CFT},q^{\alpha,\CFT}(x)]=\partial_x m^{\alpha,\CFT}(x),
\end{equation}
which on the lattice transforms to Eq.~\eqref{eq:qm} below. Then analogous to Eqs.~\eqref{eq:qCFT}-\eqref{eq:mCFT}, we can write down the lattice operators that correspond to the chiral current density $J^{\alpha}_j$ and anti-chiral current density $\bar{J}^{\alpha}_j$  as in Eq.~\eqref{eq:latticeJi} below. Finally the lattice Kac--Moody generators $J^{\alpha}_n$ and $\bar{J}^{\alpha}_n$ can be constructed as the Fourier modes of $J^{\alpha}_j$ and $\bar{J}^{\alpha}_j$. 

\begin{empheq}[box=\fbox]{align}
\label{eq:qm}
&i[H, q_j^{\alpha}] = m^{\alpha}_{j+1} - m^{\alpha}_j\\ 
\label{eq:latticeJi}
&J^{\alpha}_j = \frac{q^{\alpha}_j + m^{\alpha}_j}{2} , \quad \bar{J}^{\alpha}_j = \frac{q^{\alpha}_j - m^{\alpha}_j}{2}\\
\label{eq:J}
&J^{\alpha}_n = \sum_{j}^{N} e^{i j n \frac{2\pi}{N}} J^{\alpha}_{j}, \quad 
\bar{J}^{\alpha}_n = \sum_{j}^{N} e^{- i j n \frac{2\pi}{N}}  \bar{J}^{\alpha}_{j}.
\end{empheq}
This completes the construction of lattice Kac--Moody generators for general symmetry $G$.

\subsection{Example: XXX model with next-to-nearest neighbor coupling}
 
Consider an antiferromagnetic spin-1/2 chain with next-to-nearest neighbor coupling $J_c=0.241167$ \cite{affleck1989, eggert1996}
\begin{equation}
H =  \mathcal{N} \left( \sum_{j=1}^{N} \vec{S}_j \cdot \vec{S}_{j+1} + J_c \sum_{j=1}^N \vec{S_j} \cdot \vec{S}_{j+2} \right)
\end{equation}
where $S^{\alpha}_j = \frac{1}{2} \sigma_{j}^{\alpha}$ and $\alpha = x, y, z.$
Overall normalization factor $\mathcal{N} = 0.856 $ is determined by fixing the energy spectrum to be  Eq.~\eqref{eq:ECFT} in the large-$N$ limit. 

This model has three exact conserved charges $Q^\alpha = \sum_{j=1}^{N} q^\alpha_j = \sum_{j=1}^{N} \sqrt{2}  S_{j}^{\alpha} $ which commute with the Hamiltonian: $[Q^{\alpha}, H] = 0$. These charges satisfy $[Q^{\alpha}, Q^{\beta}] = i f^{\alpha \beta \gamma} Q^{\gamma}$ with structure constants $ f^{\alpha \beta \gamma} = \sqrt{2} \epsilon^{\alpha \beta \gamma} $. In the continuum limit, this model corresponds to the $SU(2)$ WZW model at level $k=1$. 


To find the lattice Kac--Moody generators, first we calculate the current $m^\alpha_j$ using Eq.~\eqref{eq:3local}
\begin{equation}
m_j^\alpha = - f^{\alpha \beta \gamma} \mathcal{N} \left( S_j^{\beta} S_{j+1}^{\gamma}  + J_c S_j^{\beta} S_{j+2}^{\gamma}  \right).
\end{equation}

Then following Eq.~\eqref{eq:J}, we construct $J_n^{\alpha}$ and $\bar{J}_n^{\alpha}$ as follows:

\begin{equation}
\begin{aligned}
J_n^{\alpha} =  \sum_{j=1}^{N_0} e^{+ijn \frac{2\pi}{N}}  \frac{\sqrt{2}}{2}
\left[
S_j^\alpha - \epsilon^{\alpha \beta \gamma} \mathcal{N}  \left( S_j^{\beta} S_{j+1}^{\gamma}  +  J_c S_j^{\beta} S_{j+2}^{\gamma}  \right) 
\right], \\
\bar{J}_n^{\alpha} =  \sum_{j=1}^{N_0} e^{-ijn \frac{2\pi}{N}}  \frac{\sqrt{2}}{2}
\left[
S_j^\alpha + \epsilon^{\alpha \beta \gamma} \mathcal{N}  \left( S_j^{\beta} S_{j+1}^{\gamma}  +  J_c S_j^{\beta} S_{j+2}^{\gamma}  \right) 
\right].
\end{aligned}
\end{equation}

In order to confirm that they correspond to Kac--Moody generators in the continuum, we perform simultaneous exact diagonalization of the above lattice Hamiltonian $H$ and of the lattice translation operator to obtain the low-energy eigenstates. Scaling dimensions and conformal spins can be computed using Eqs.~\eqref{eq:Elat}-\eqref{eq:Plat}. 

\begin{figure}
\centering
\includegraphics[width=\linewidth]{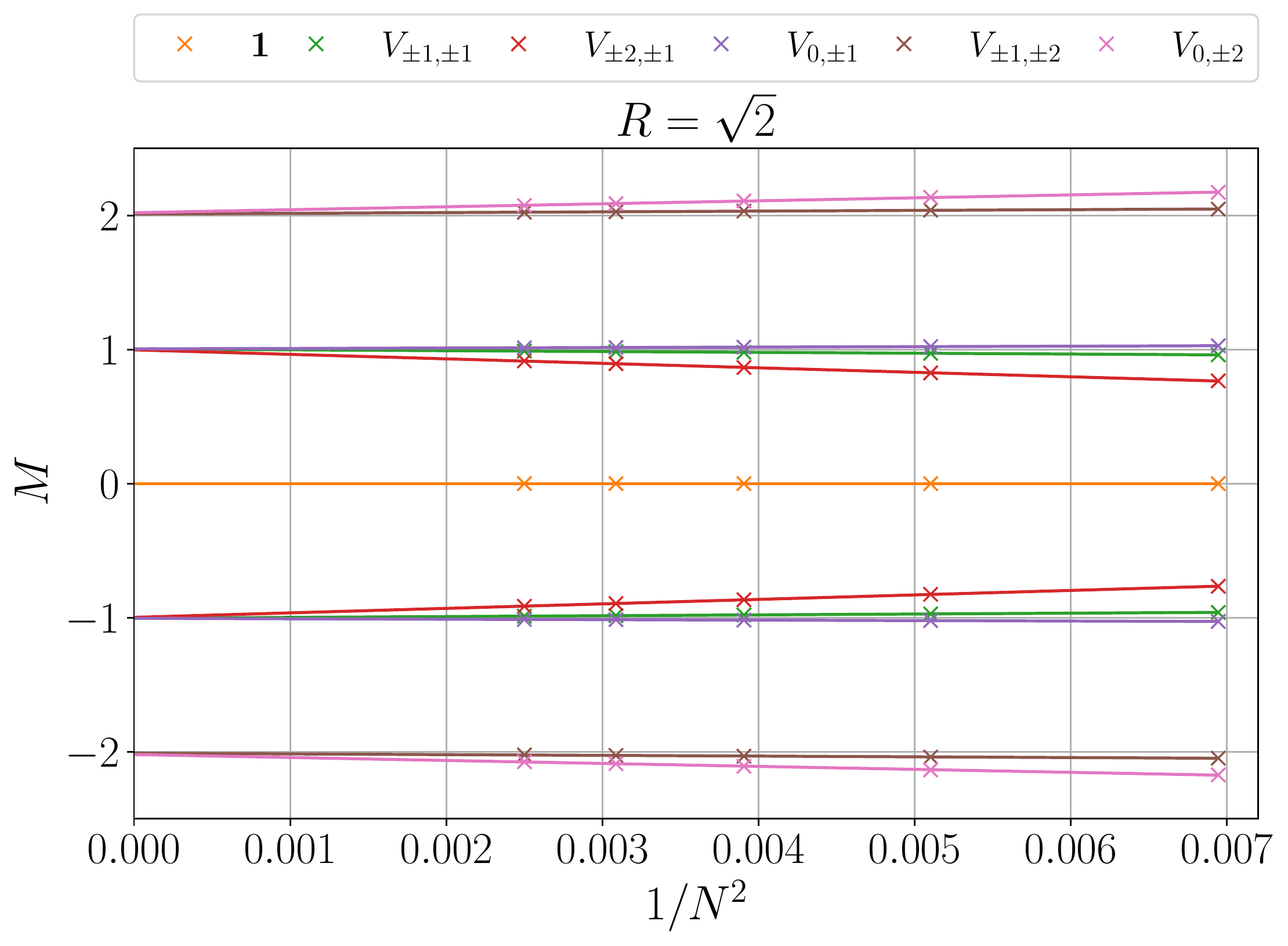}
\caption{Expectation values of $M^z$ approach integers in the thermodynamic limit. Finite size scaling of expectation value $\mel{\psi}{M^z}{\psi}$, for $\ket{\psi} = \ket{\mathbf{1}}, \ket{V_{\pm1, \pm1}}, \ket{V_{\pm2, \pm1}}, \ket{V_{0, \pm1}}, \ket{V_{\pm1, \pm2}}, \ket{V_{0, \pm2}}$, extrapolated from system sizes $N=12, 14, 16, 18, 20$. This suggests that $M^z$ is an emergent conserved charge.}
\label{fig:xxx_m}
\end{figure}

Now we check various matrix elements of the lattice Kac--Moody generators. We start by examining the zero modes of the lattice generators $J_0^{\alpha}$ and $\bar{J}_0^{\alpha}$. While $Q^{\alpha} = J_0^{\alpha} + \bar{J}_0^{\alpha}$ are exact symmetries, the charges $M^{\alpha} = J_0^{\alpha} - \bar{J}_0^{\alpha}$ do not commute with the Hamiltonian on the lattice, i.e. $[M^{\alpha}, H] \neq 0$. 
However linear extrapolation of matrix elements $\mel{\psi}{M^\alpha}{\psi}$ suggests that in the thermodynamic limit each $M^\alpha$ becomes indeed a conserved charge, as illustrated in Fig.~\ref{fig:xxx_m}. We also numerically confirmed that the matrix elements $\mel{\psi}{[Q^\alpha, M^\beta]}{\psi}$ approximately vanish for low energy eigenstates, up to finite size error (not shown in Fig.~\ref{fig:xxx_m}).

\begin{figure}
\centering
\includegraphics[width=\linewidth]{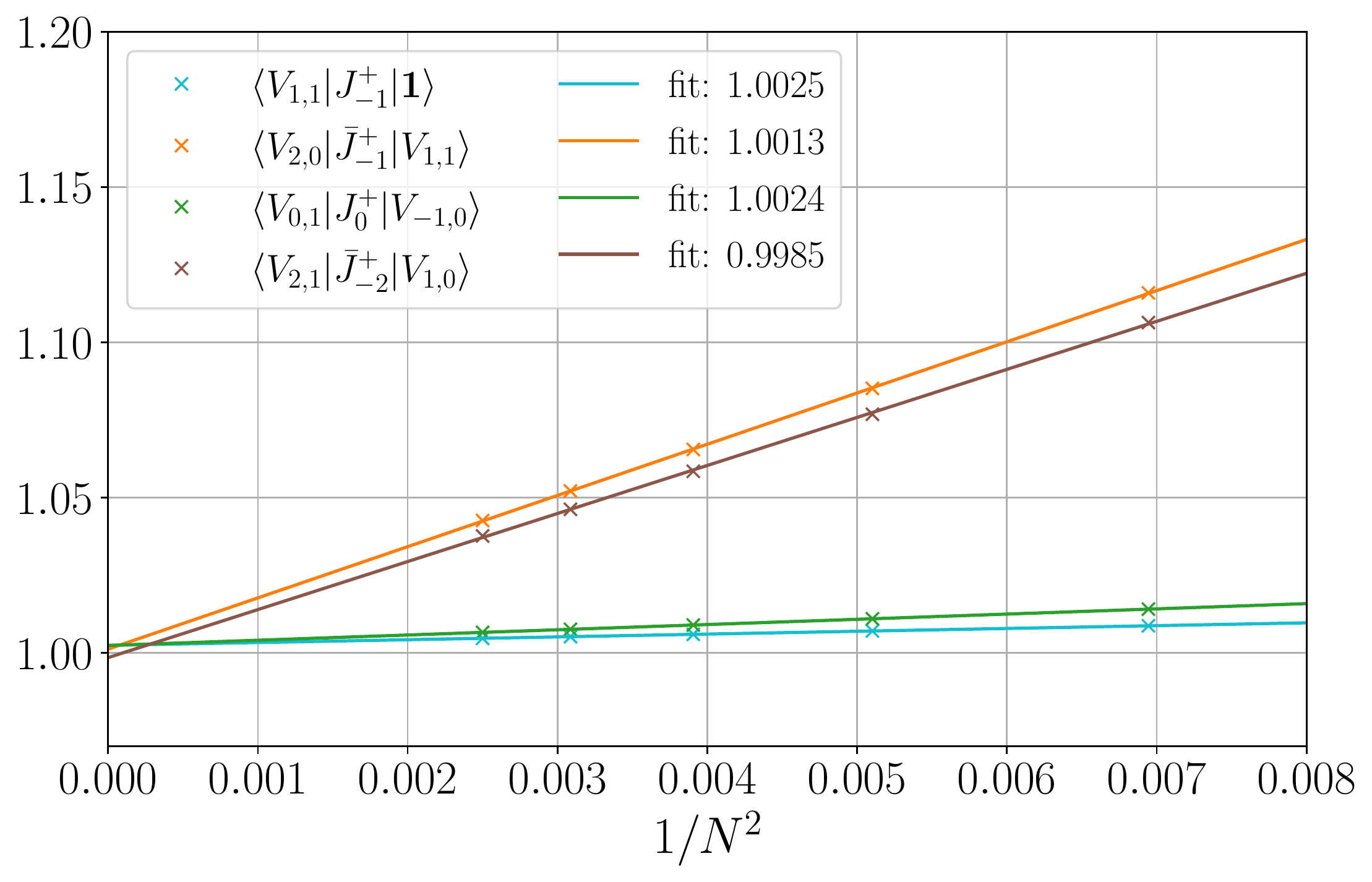}
\caption{ Matrix elements from Eqs.~\eqref{eq:kmtower1}-\eqref{eq:kmtower3} and Eq.~\eqref{eq:kmtower5}.}
\label{fig:xxx_matele}
\end{figure}

Kac--Moody primary states can be identified numerically by checking conditions in Eq.~\eqref{eq:hws}. Up to finite size error, the only eigenstates that satisfy the conditions in Eq.~\eqref{eq:hws} are $\ket{\mathbf{1}}=\ket{V_{0,0}}$ and $\ket{V_{0, 1}}$. For other Virasoro primary states that are not Kac--Moody primary states, we confirm that they can be obtained as descendants of these two Kac--Moody primary states using the lattice generators. For example, among the four Virasoro primary states $\ket{V_{0, \pm1}}$ and $\ket{V_{\pm 1, 0}}$ at conformal spin $S=0$ and dimension $\Delta=1/2$, only $\ket{V_{1, 0}}$ is a Kac--Moody primary state. We show that $\ket{V_{0, \pm1}}$ are indeed Kac--Moody descendent states of $\ket{V_{1, 0}}$ by checking matrix elements $\mel{V_{0, \pm1}}{J^{\pm}_{0}}{V_{1, 0}}$, as plotted in Fig.~\ref{fig:xxx_matele}. 

\begin{figure}
\begin{minipage}[b]{0.78\linewidth}
\centering
\includegraphics[width=\textwidth]{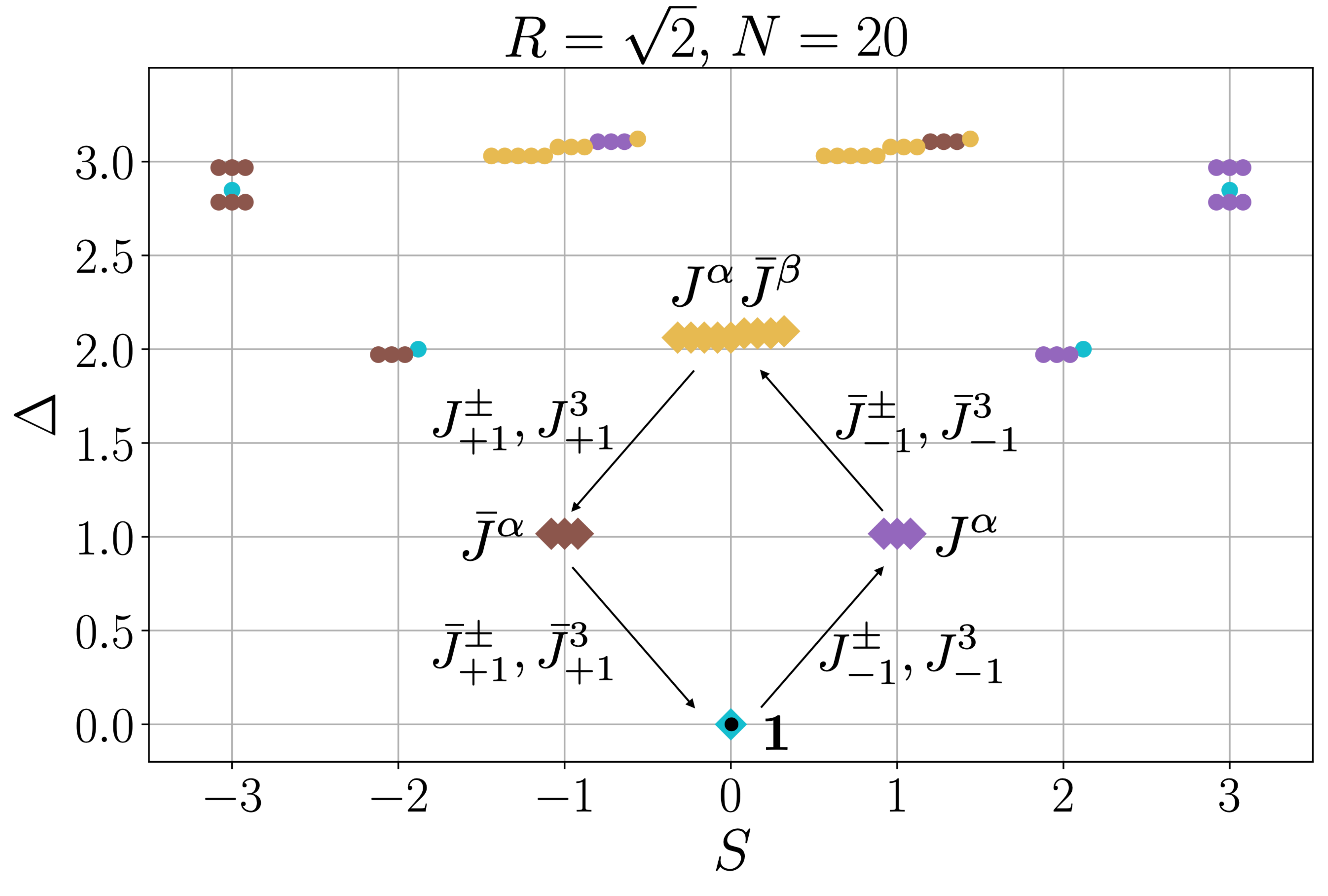}
\end{minipage}
\begin{minipage}[b]{0.78\linewidth}
\centering
\includegraphics[width=\textwidth]{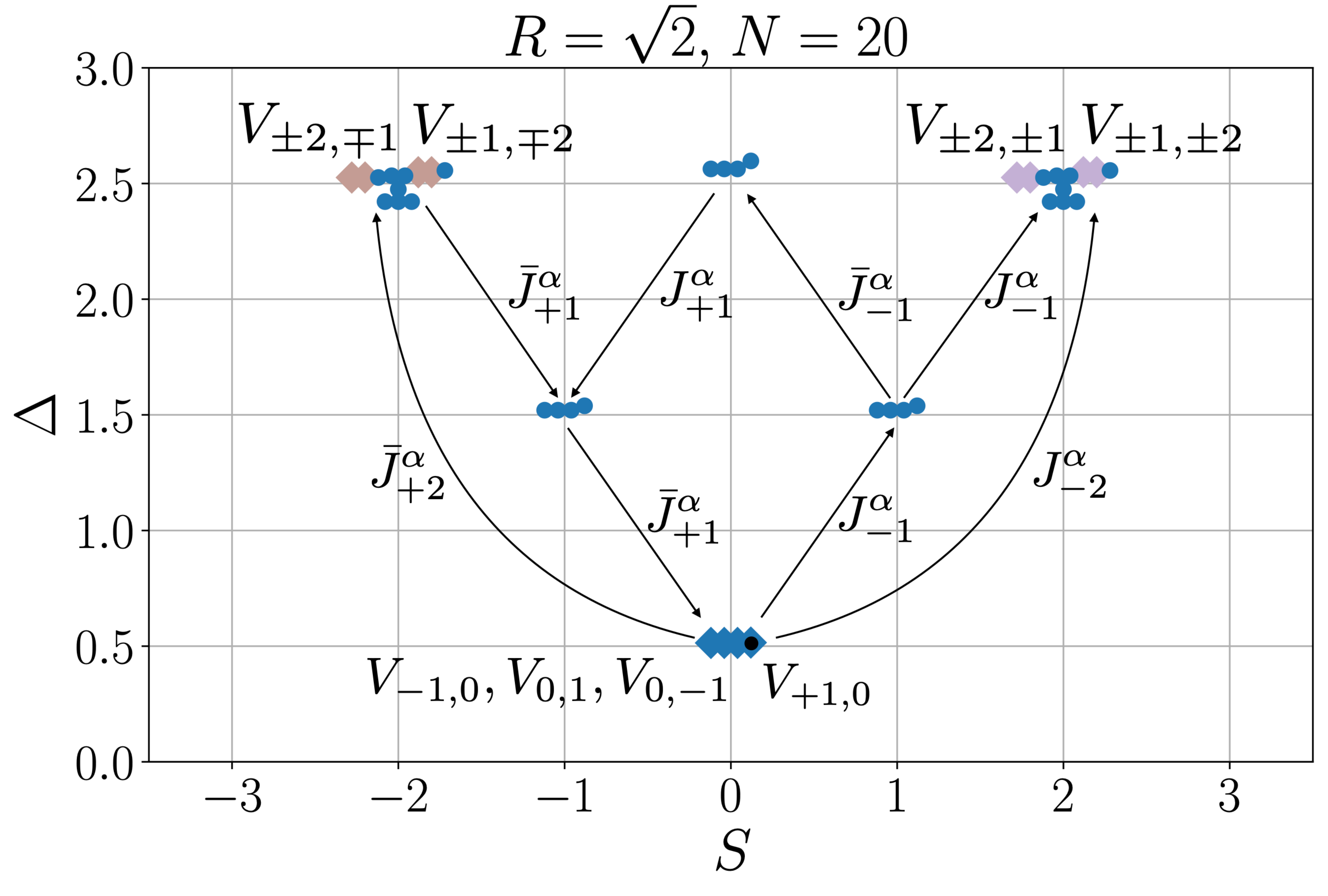}
\end{minipage}
\caption{XXX model spectrum with NNN interaction at system size $N = 20$, separated into two Kac--Moody towers. Top: Kac--Moody tower of $\ket{\mathbf{1}}$. Bottom: Kac--Moody tower of $\ket{V_{1,0}}$. Some examples of actions of $J^{\alpha}_n$ and $\bar{J}^{\alpha}_n$ are plotted.}
\label{fig:spectrum_xxx}
\end{figure}

Acting on the Kac--Moody primary states with $J_{-n}^\alpha$ and $\bar{J}_{-n}^\alpha$, all other descendent states can be reached and classified into Kac--Moody towers. While $J^3_n$ preserves the eigenvalue of $J^z$,  $J^{\pm}_{-n}$ changes the eigenvalue of $J^z$ by $ \pm n$. We illustrate some examples of the actions of $J^{\alpha}_{-n}$ and $\bar{J}^{\alpha}_{-n}$ in Fig.~\ref{fig:spectrum_xxx} and present the finite size scaling results of some corresponding matrix elements in Fig.~\ref{fig:xxx_matele}.

\begin{figure}
\begin{minipage}[b]{0.49\linewidth}
\centering
\includegraphics[width=\textwidth]{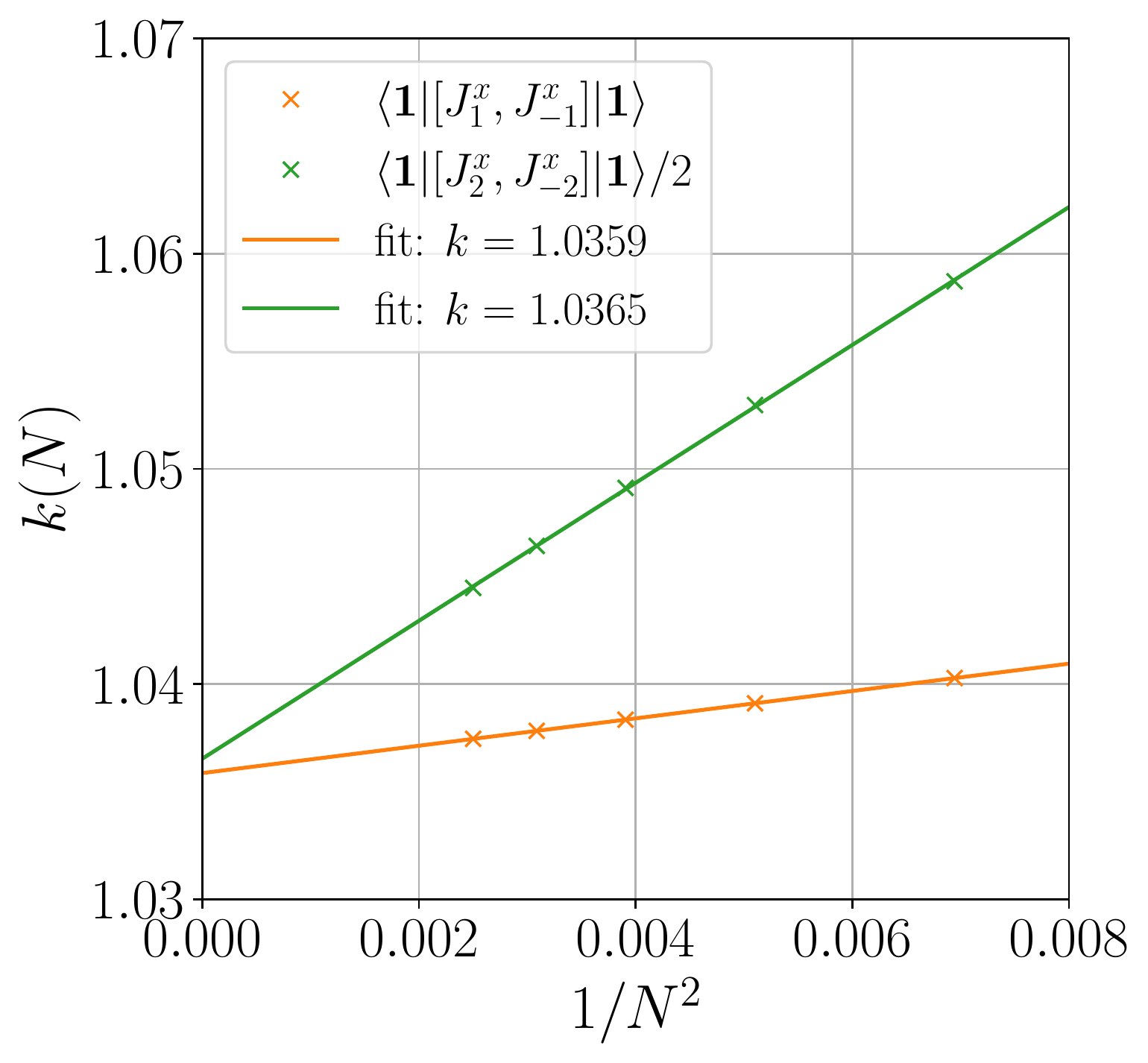}
\end{minipage}
\begin{minipage}[b]{0.46\linewidth}
\centering
\includegraphics[width=\textwidth]{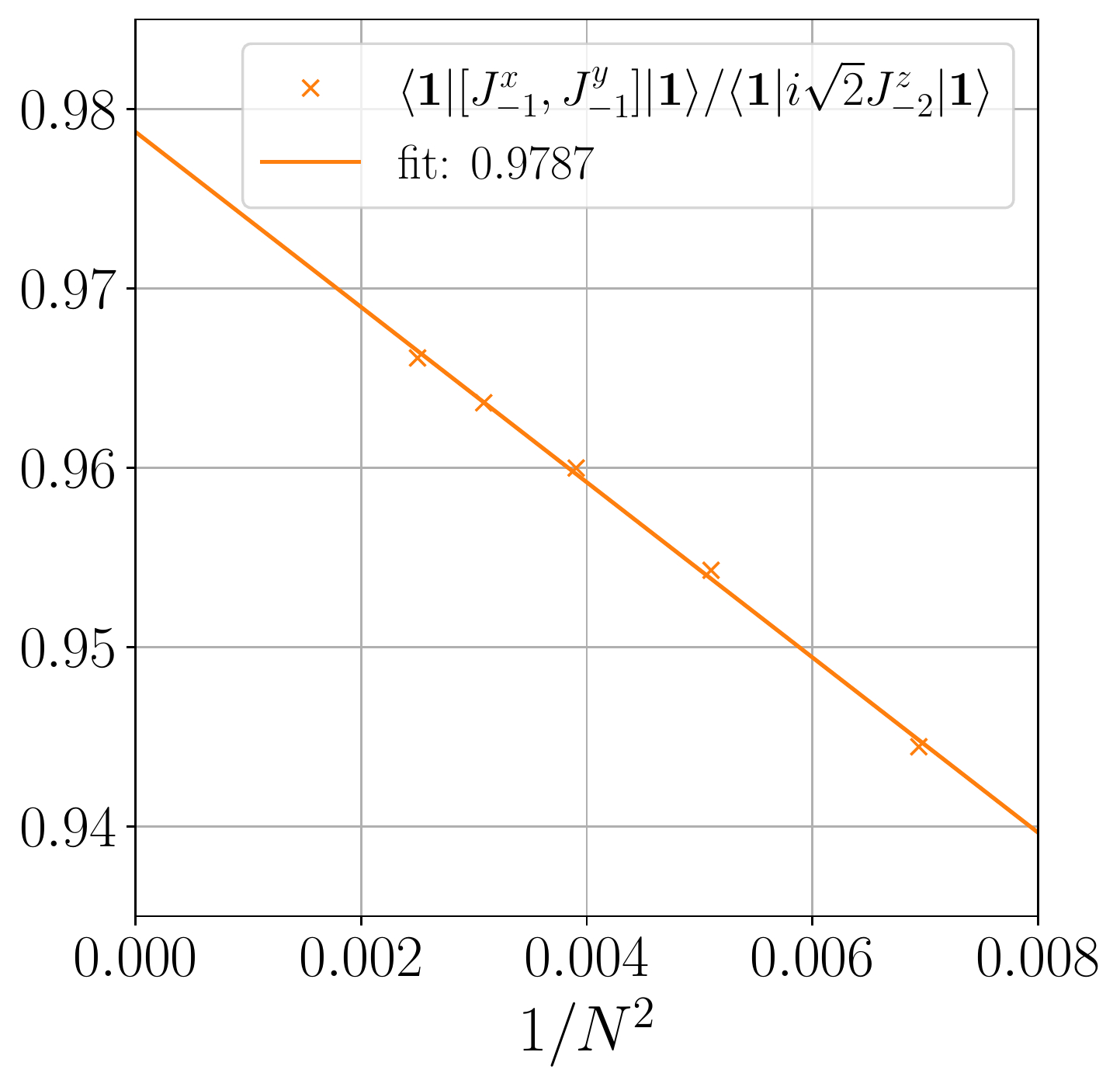}
\end{minipage}
\caption{ Left: Level constant extracted from matrix elements $\mel{\mathbf{1}}{[J^x_{+1}, J^x_{-1}]}{\mathbf{1}}$ and $\mel{\mathbf{1}}{[J^x_{+2}, J^x_{-2}]}{\mathbf{1}}$. Right: Finite size scaling of structure constant $f$ from  Eq.~\eqref{eq:struct}.}
\label{fig:xxx_kc}
\end{figure}

Moreover, we are able to confirm not only the level constant $k$ but also the structure constants $f^{\alpha \beta \gamma}$. While the structure constants of the diagonal subalgebra are given, it is non-trivial to check the commutation relations of the chiral algebra Eq.~\eqref{eq:struct}. For example, we confirm $[J^x_{m}, J^y_{n}]=i \sqrt{2}J^z_{m+n}$ in the low energy subspace by calculating the ratio between $\mel{\psi}{[J^x_{m}, J^y_{n}]}{\psi}$ and $\mel{\psi}{i \sqrt{2}  J^z_{m+n}}{\psi}$. In Fig.~\ref{fig:xxx_kc}, we show that in the thermodynamic limit $\mel{\mathbf{1}}{J^y_{-1} J^x_{-1}}{\mathbf{1}} \approx 0.9787 \times \left(i \sqrt{2} \mel{\mathbf{1}}{J^z_{-2}}{\mathbf{1}} \right)$, which is in good agreement with Eq.~\eqref{eq:struct}, up to finite-size effects. Other components of $f^{\alpha \beta \gamma}$ can be checked numerically in similar fashion using the proposed lattice Kac--Moody generators.

\section{Discussion}
\label{sec:discussion}
Since the groundbreaking work by Cardy and others in the 1980s, there have been a series of studies on numerically obtaining conformal data from critical lattice systems (see e.g. \cite{cardy1984,blote_1986,affleck1986,cardy1986, ks, READ2007316, DUBAIL2010399, Gainutdinov_2013, GAINUTDINOV2013223, Bondesan:2014hza, mv, zmv1, zmv2, zv}). Our paper adds new capabilities and applications to this line of work, by addressing the very important case of quantum critical spin chains with a global symmetry.

As we have reviewed, a critical quantum spin chains with a microscopic Lie group symmetry $G$ corresponds to a CFT with Kac--Moody symmetry. In this paper, we have proposed a concrete construction of Kac--Moody currents $J_n^{\alpha}$ and $\bar{J}_n^{\alpha}$ on lattice with the Hamiltonian and its microscopic symmetry as the only input. Our construction allows us to observe the emergence of Kac--Moody symmetries numerically. 

We illustrated our methods using two spin models, each having a global symmetry with Lie group $G$, namely the XXZ model as an Abelian example and the XXX (or Heisenberg) model with next-to-nearest neighbor coupling as a non-Abelian example. First, we obtained the low-energy spectrum of the Hamiltonian using exact diagonalization. We then verified our proposal by studying the action of lattice Kac--Moody generators on low-energy energy-momentum eigenstates. As we consider larger system sizes, a second copy of the symmetry is observed to emerge, making the global symmetry $G\times G$. We also successfully demonstrated the identification of the Kac--Moody primary states and their Kac--Moody towers. We show that the proposal works for both the Abelian and non-Abelian cases, and with independence of whether the model is integrable. 

To apply our method to larger system sizes, and thus further reduce finite-size corrections in the extracted conformal data, tensor network techniques such as puMPS \cite{zmv1} can be employed. It would also be interesting to study the realizations of other extended symmetries on the lattice, such as the $\mathcal{W}$-algebra appearing in the three-state Potts model \cite{FATEEV1987644}. 

\textit{Note:} In subsequent work (which was posted recently \cite{wang2022}), one of the authors generalized some of the aspects of our work to a more general setting that includes critical systems for which an explicit local Hamiltonian description is not available.

\section{Acknowledgement}
R.W. acknowledges support from the Perimeter Institute for Theoretical Physic through the Visiting Graduate Fellowship Program where part of this research was done. Y.Z. is supported by the Q-FARM fellowship at Stanford University. G.V. is a CIFAR fellow in the Quantum Information Science Program, a Distinguished Invited Professor at the Institute of Photonic Sciences (ICFO), and a Distinguished Visiting Research Chair at Perimeter Institute. Research at Perimeter Institute is supported in part by the Government of Canada through the Department of Innovation, Science and Economic Development Canada and by the Province of Ontario through the Ministry of Colleges and Universities.


\bibliography{reference}

\newpage
\appendix
\section{Continuum limit of XX model and Abelian bosonization}
\label{appendix:a}

In this appendix, we discuss lattice Kac--Moody generator construction of XX model through bosonization \cite{fradkin2013}. The result is consistent with the Kac--Moody generator we found in the main text.

The Hamiltonian of XX model on a 1-dimensional lattice of $N$ sites is
\begin{equation}
H = \sum_{k=1}^{N} \left(
S^X_k S^X_{k+1} + S^Y_k S^Y_{k+1}
\right)
\end{equation}
where $S^\alpha = \sigma^\alpha/2$.
It can be brought to a fermionic representation
\begin{equation}
H = \frac{1}{2} \sum_{k=1}^{N} \left(
i a^{\dagger}_k a_{k+1} + \text{h.c.}
\right)
\label{eq:fermiH}
\end{equation}
by introducing fermionic variables 
\begin{equation}
a_k = i^{-k} \exp(\pi i \sum_{j=1}^{k-1} S_j^+ S_j^-) S^-_k 
\end{equation}
where $S_k^{\pm} =S^X_k \pm i S^Y_k$.

Define spinor field $\phi_\alpha(\alpha = 1, 2)$ by 
\begin{equation}
\phi_{\alpha}(k) =
\begin{cases}
\phi_1(k) = a_{2s} \quad\quad k=2s\\
\phi_2(k) = a_{2s+1} \quad k=2s+1.
\end{cases}
\end{equation}
It could be easily checked that 
\begin{equation}
\{\phi^{\dagger}_{\alpha}(n), \phi_{\beta}(m)\} = \delta_{\alpha \beta} \delta_{mn}.
\end{equation}

In the continuum limit, define  $\psi_\alpha(x) = \frac{1}{\sqrt{2\Lambda}} \phi_\alpha (k)$ where $x=2 s \Lambda$ and $\Lambda$ is the lattice spacing. Then Eq.~\eqref{eq:fermiH} can be rewritten as 
\begin{equation}
\begin{aligned}
H & = \frac{i}{2}  \sum_{k=1}^{N} a^\dagger_k (a_{k+1} - a_{k-1})\\
& =  \frac{i}{2}  \sum_{s=1}^{N/2} \{ \phi_1^\dagger(2s) (\phi_2(2s+1)-\phi_2(2s-1)) \\
& \qquad \qquad + \phi_1^\dagger(2s+1) (\phi_2(2s+2)-\phi_2(2s)) \} \\
& \sim \int_{0}^{N\Lambda} dx \, \psi^\dagger(x) \sigma^x i \partial_x \psi(x),\\
\end{aligned}
\end{equation}
where we have replaced finite differences by derivatives and the finite sum by an integral.
The Hamiltonian has a $U(1)$ symmetry $\psi(x)\rightarrow e^{i\alpha}\psi(x)$. Fermionic currents $j_\mu$ associated with the symmetry are 
\begin{equation}
\begin{aligned}
&j_0 = \psi_1^\dagger \psi_1 + \psi_2^\dagger \psi_2 \\
&j_1 = \psi_2^\dagger \psi_1 - \psi_1^\dagger \psi_2
\end{aligned}
\end{equation}
where $j_0$ is the total charge density and $j_1$ equals the difference of densities of left and right movers. To make connection with our notation, we can go back to the lattice where the currents can be written as 
\begin{equation}
\begin{aligned}
&j_0 \sim S^X_{k} S^X_{k+1} + S^Y_{k} S^Y_{k+1} \\
&j_1 \sim S^X_{k} S^Y_{k+1} - S^Y_{k} S^X_{k+1}.
\end{aligned}
\end{equation}
Currents $j_0$ and $j_1$ correspond to $q$ and $m$ in our notation.

\section{Matrix elements of Kac--Moody generators}
\label{appendix:b}

In this appendix we compute matrix elements of Kac--Moody generators that appear in the main text. We will omit the $^\CFT$ superscript in the appendix. It should be understood that all operators and states are in the CFT. Here we will only discuss the chiral part of the operators, the anti-chiral part yields analogous properties.

First, consider the $\mathfrak{u}(1)_k$ Kac--Moody algebra. We would like to show that
\begin{equation}
\label{Jngr}
    J_{-n}|\mathbf{1}\rangle=\sqrt{nk} |\partial^{n-1}J\rangle ~ (n\geq 1).
\end{equation}
On the cylinder $J_n$ is the Fourier mode of the current density $J(x)$,
\begin{equation}
    J_{-n}=\frac{1}{2\pi}\int_0^{L} dx\, e^{-inx\frac{2\pi}{L}} J(x).
\end{equation}
This can be mapped to the complex plane, where the integral in $x$ becomes a contour integral around the origin.
\begin{equation}
    J_{-n}=\frac{1}{2\pi i}\oint_0 dz z^{-n} J(z).
\end{equation}
Using the residue theorem, when $n\geq 1$, 
\begin{equation}
    J_{-n}= (n-1)!\frac{d^{n-1}}{dz^{n-1}} J(z)|_{z=0}.
\end{equation}
Using the operator-state correspondence,
\begin{equation}
    O(0)|\mathbf{1}\rangle \propto |O\rangle,
\end{equation}
we know that $J_{-n}$ acting on the vacuum gives the $|\partial^{n-1}J\rangle$ state. The overall normalization $\sqrt{nk}$ can be determined by 
\begin{equation}
    \langle\mathbf{1}|J_nJ_{-n}|\mathbf{1}\rangle= nk ~ (n\geq 1),
\end{equation}
which follows from the Kac--Moody algebra.

Next, we compute the action of $J_{n}$ on a vertex operator state $|V_{Q,M}\rangle$. To do this, first recall that the vertex operator can be decomposed into a chiral part and an anti-chiral part,
\begin{equation}
    V_{Q,M}(z,\bar{z})=e^{i\alpha \phi_L}(z)e^{i\beta\phi_R}(\bar{z}),
\end{equation}
where $\phi_L$ and $\phi_R$ represent the chiral and anti-chiral boson operator, and
\begin{eqnarray}
\label{alphaR}
  \alpha &=& \frac{Q}{R}+\frac{MR}{2}\\
  \label{betaR}
  \beta &=& \frac{Q}{R}-\frac{MR}{2},
\end{eqnarray}
where $R$ is the compactification radius. The Kac--Moody generators $J_{n}$ only act on the chiral part $e^{i\alpha \phi_L}$. If $\alpha=0$, then the chiral part is the identity operator. Then 
\begin{equation}
    J_{-n}|V_{Q,M}\rangle=\sqrt{nk}|\partial^{n-1}J e^{i\beta\phi_R}\rangle ~ (n\geq 1),
\end{equation}
where $\partial^{n-1}J e^{i\beta\phi_R}$ is the composite operator of the chiral component $\partial^{n-1}J$ and the anti-chiral component $e^{i\beta\phi_R}$. This is a result of combining Eq.~\eqref{Jngr} and the anti-chiral part of the vertex operator. If $n<1$, then $J_{-n}$ annihilates the vertex operator state. Eq.~\eqref{Jngr} can be seen as a special case where $Q=M=0$.

Below we consider the case where $\alpha\neq 0$. The result of acting $J_n$ on a vertex operator can be represented as the contour integral of an operator product expansion (OPE),
\begin{equation}
    (J_n e^{i\alpha\phi_L})(w)=\frac{1}{2\pi i}\oint_z dz\, (z-w)^n J(z)e^{i\alpha\phi_L}(w),
\end{equation}
where
\begin{equation}
    J(z)=i\partial\phi(z).
\end{equation}
The OPE can be computed in a standard way via Wick contraction,
\begin{equation}
    J(z)e^{i\alpha\phi_L}(w)=\frac{\alpha}{z-w}e^{i\alpha\phi_L}(w)+\frac{1}{\alpha} \partial_w e^{i\alpha\phi_L}(w)+\cdots,
\end{equation}
where $\cdots$ represents terms with positive powers of $z-w$. This implies that
\begin{eqnarray}
(J_n e^{i\alpha\phi_L})(w) &=& 0 ~ (n>0) \\
(J_0 e^{i\alpha\phi_L})(w) &=& \alpha e^{i\alpha\phi_L}(w) \\
(J_{-1} e^{i\alpha\phi_L})(w) &=& \frac{1}{\alpha}\partial_w e^{i\alpha\phi_L}(w).
\end{eqnarray}
The first equation implies that the vertex operator is a Kac--Moody primary operator. The second equation implies (via operator-state correspondence) that $|V_{Q,M}\rangle$ is an eigenstate of $J_0$,
\begin{equation}
    J_0|V_{Q,M}\rangle= \alpha |V_{Q,M}\rangle.
\end{equation}
The third equation implies 
\begin{equation}
    J_{-1}|V_{Q,M}\rangle= \sqrt{k}|\partial V_{Q,M}\rangle,
\end{equation}
where we have normalized the state $|\partial V_{Q,M}\rangle$ to unit norm. In general, acting with $J_{-n}$ on a vertex operator produces a linear combination of Virasoro descendant states at level $n$.

Next, we consider the $\mathfrak{su}(2)_1$ Kac--Moody algebra. Note that $J^{3}_n$ is identical to $J_n$ in the $\mathfrak{u}(1)_1$ case with the compactification radius $R=\sqrt{2}$. We will then focus on the action of $J^{+}_n$ and $J^{-}_n$. Recall that
\begin{equation}
    J^{+}(z)=V_{1,1}(z), ~ J^{-}(z)=V_{-1,-1}(z).
\end{equation}
By the same argument that leads to Eq.~\eqref{Jngr}, we have 
\begin{equation}
    J^{\pm}_{-n}|\mathbf{1}\rangle=\sqrt{n}|\partial^{n-1}V_{\pm 1,\pm 1}\rangle.
\end{equation}
In order to see how the Kac--Moody generators act on vertex operator states, we need the OPE of vertex operators,
\begin{widetext}
\begin{equation}
   V_{Q,M}(z,\bar{z})V_{Q',M'}(w,\bar{w})= 
    (z-w)^{\alpha\alpha'}(\bar{z}-\bar{w})^{\beta\beta'}V_{Q+Q',M+M'}(w,\bar{w})+\cdots,
\end{equation}
where $\alpha,\beta$ and $\alpha',\beta'$ are determined by $Q,M$ and $Q',M'$ with Eqs.~\eqref{alphaR}\eqref{betaR}, respectively, and $\cdots$ represents terms that contain descendants of $V_{Q+Q',M+M'}$. In particular, at $R=\sqrt{2}$,
\begin{eqnarray}
J^{+}(z) V_{Q,M}(w,\bar{w})&=&(z-w)^{Q+M} V_{Q+1,M+1}(w,\bar{w})+\cdots \\
J^{-}(z) V_{Q,M}(w,\bar{w})&=&(z-w)^{-(Q+M)} V_{Q-1,M-1}(w,\bar{w})+\cdots 
\end{eqnarray}
\end{widetext}
Let $Q=1,M=0$ and $w=0$, we obtain
\begin{eqnarray}
J^{+}(z) V_{1,0}(0,0)&=&z V_{2,1}(0,0)+\cdots \\
J^{-}(z) V_{1,0}(0,0)&=&z^{-1} V_{0,-1}(0,0)+\cdots 
\end{eqnarray}
Acting with both sides on the vaccum states and using the Laurent expansion of $J^{\pm}(z)$, we obtain
\begin{eqnarray}
J^{+}_{-2} |V_{1,0}\rangle &=& |V_{2,1}\rangle \\
J^{-}_{0} |V_{1,0}\rangle&=&|V_{0,-1}\rangle.
\end{eqnarray}
Other matrix elements used in the main text can be derived in an analogous way.

\section{XXZ model at different radii}
\label{appendix:c}

\begin{figure}
\begin{minipage}{0.8\linewidth}
\centering
\includegraphics[width=\textwidth]{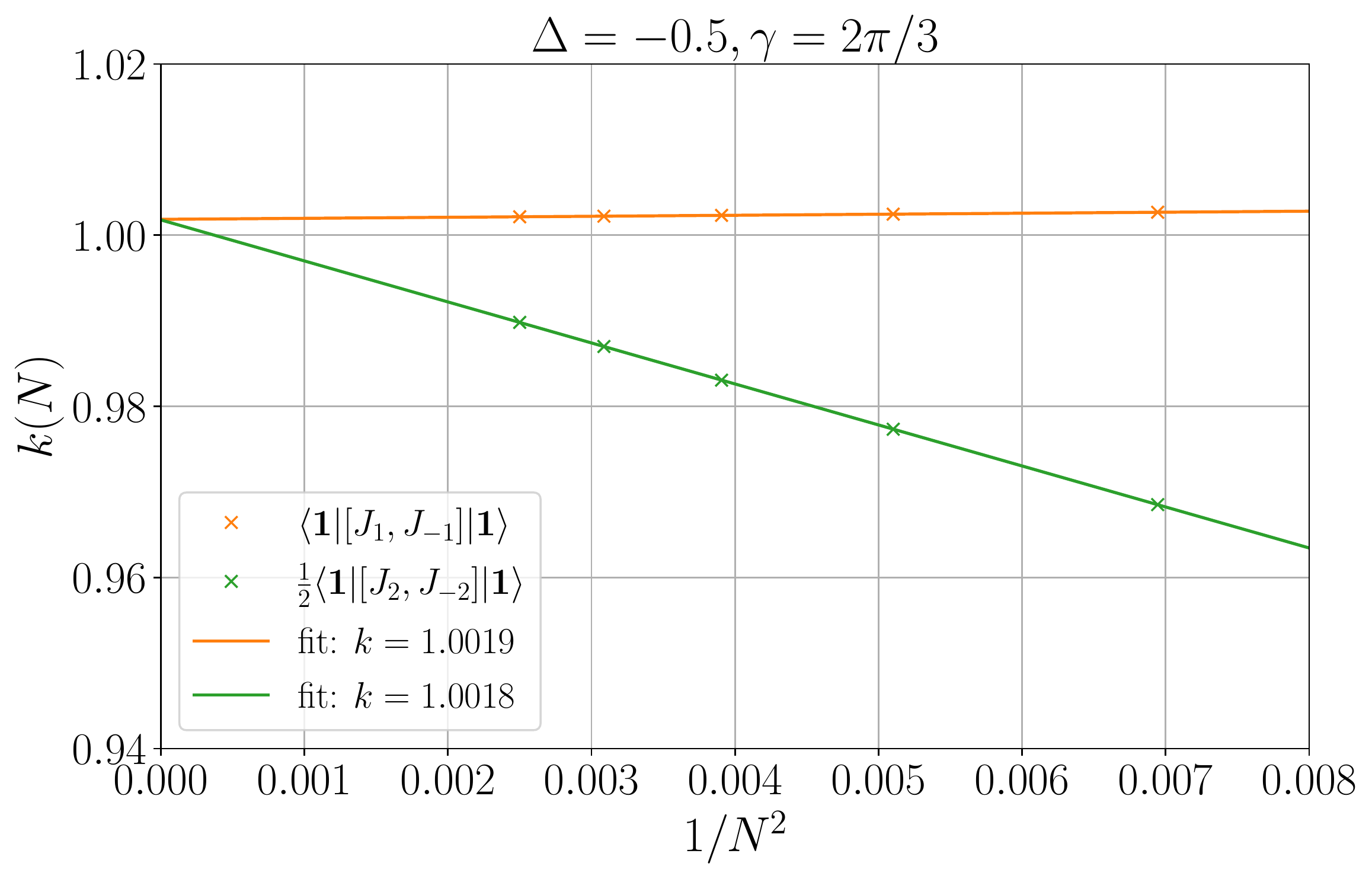}
\end{minipage}
\begin{minipage}{0.8\linewidth}
\centering
\includegraphics[width=\textwidth]{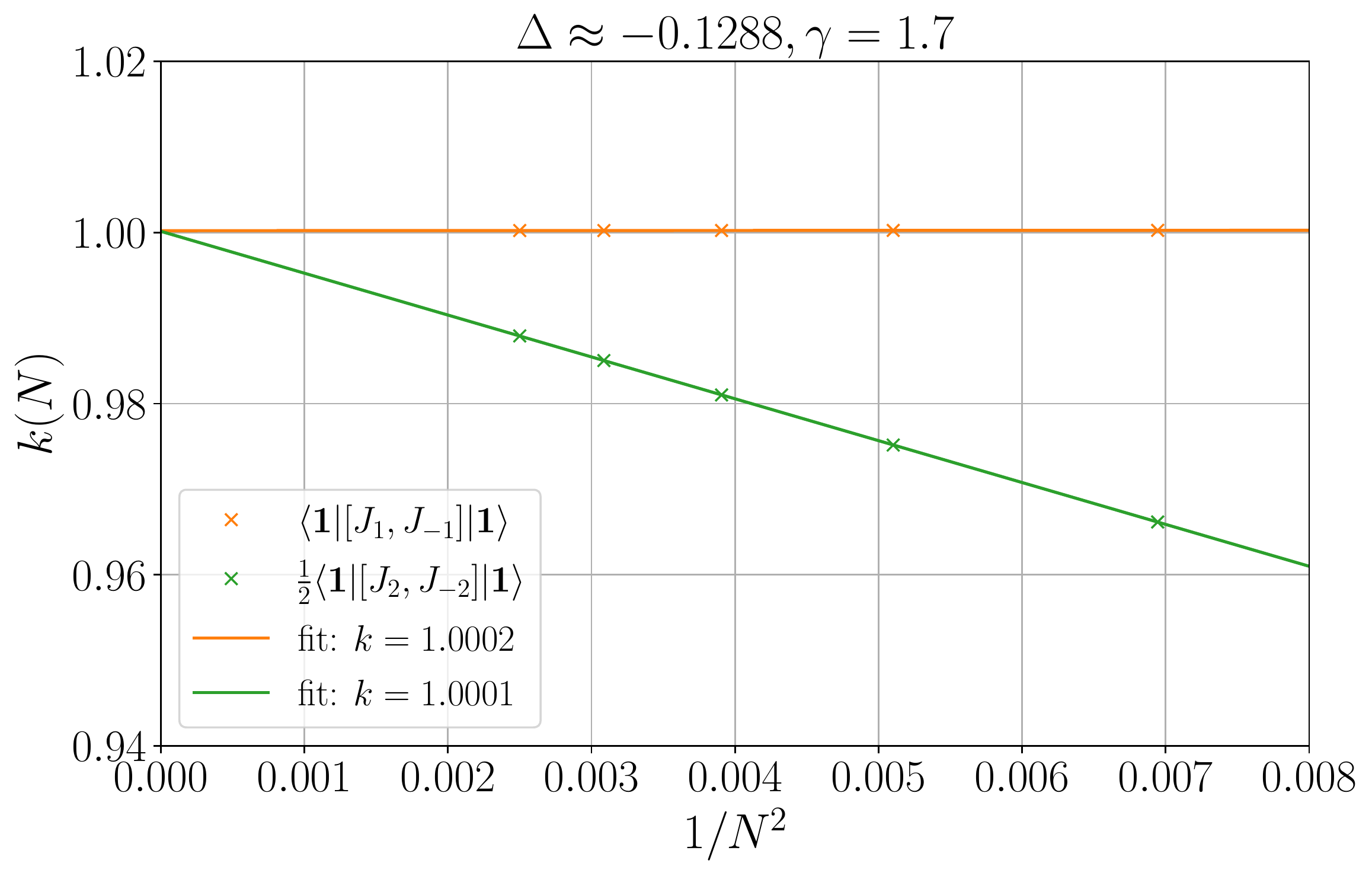}
\end{minipage}
\caption{Finite size scaling of level constant $k$ from XXZ model with $\gamma=2\pi/3$  at system sizes $N=12,14,16,18,20$.}
    \label{fig:xxz_k}
\end{figure}

\begin{figure}
\begin{minipage}{0.8\linewidth}
\centering
\includegraphics[width=\textwidth]{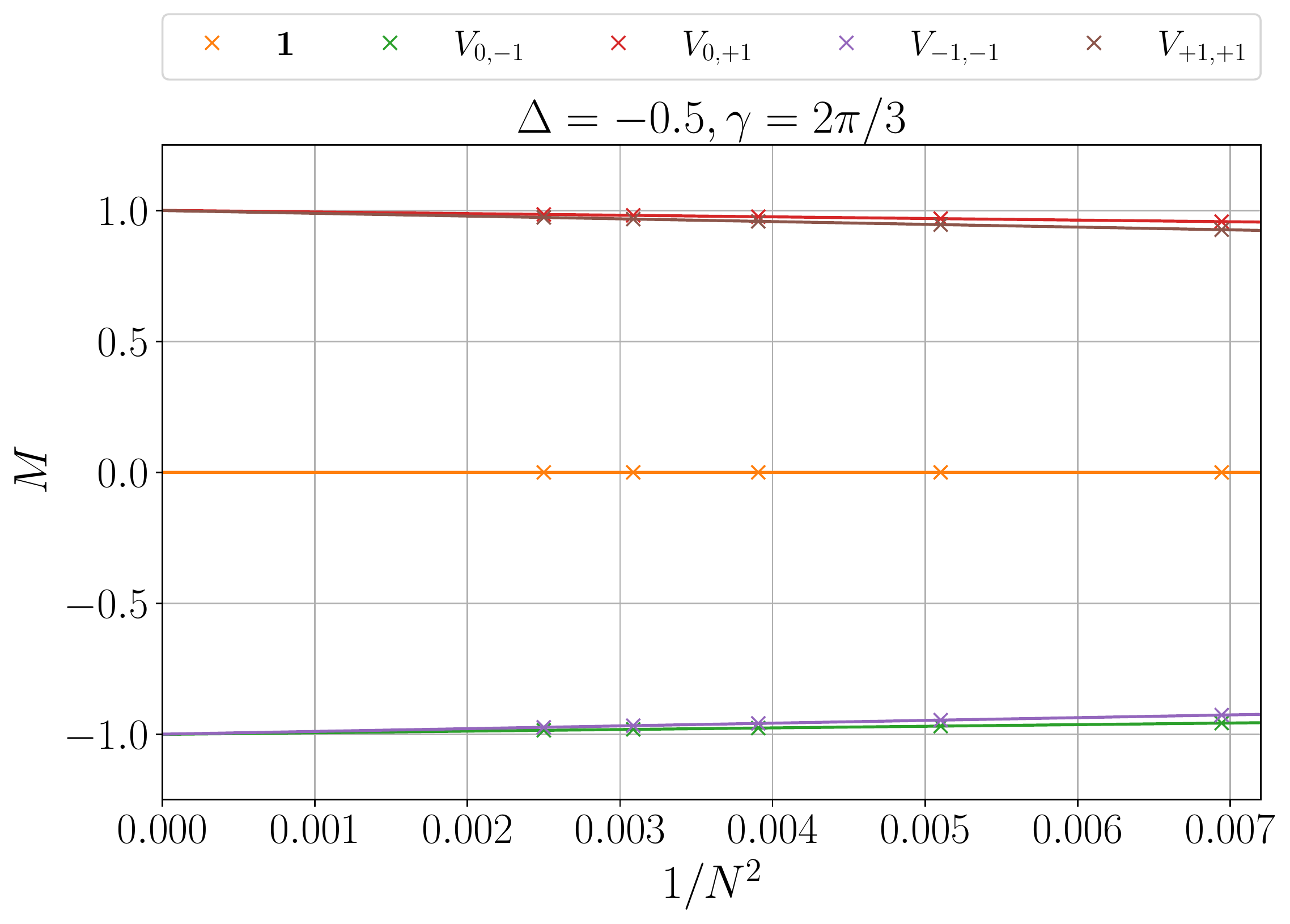}
\end{minipage}
\begin{minipage}{0.8\linewidth}
\centering
\includegraphics[width=\textwidth]{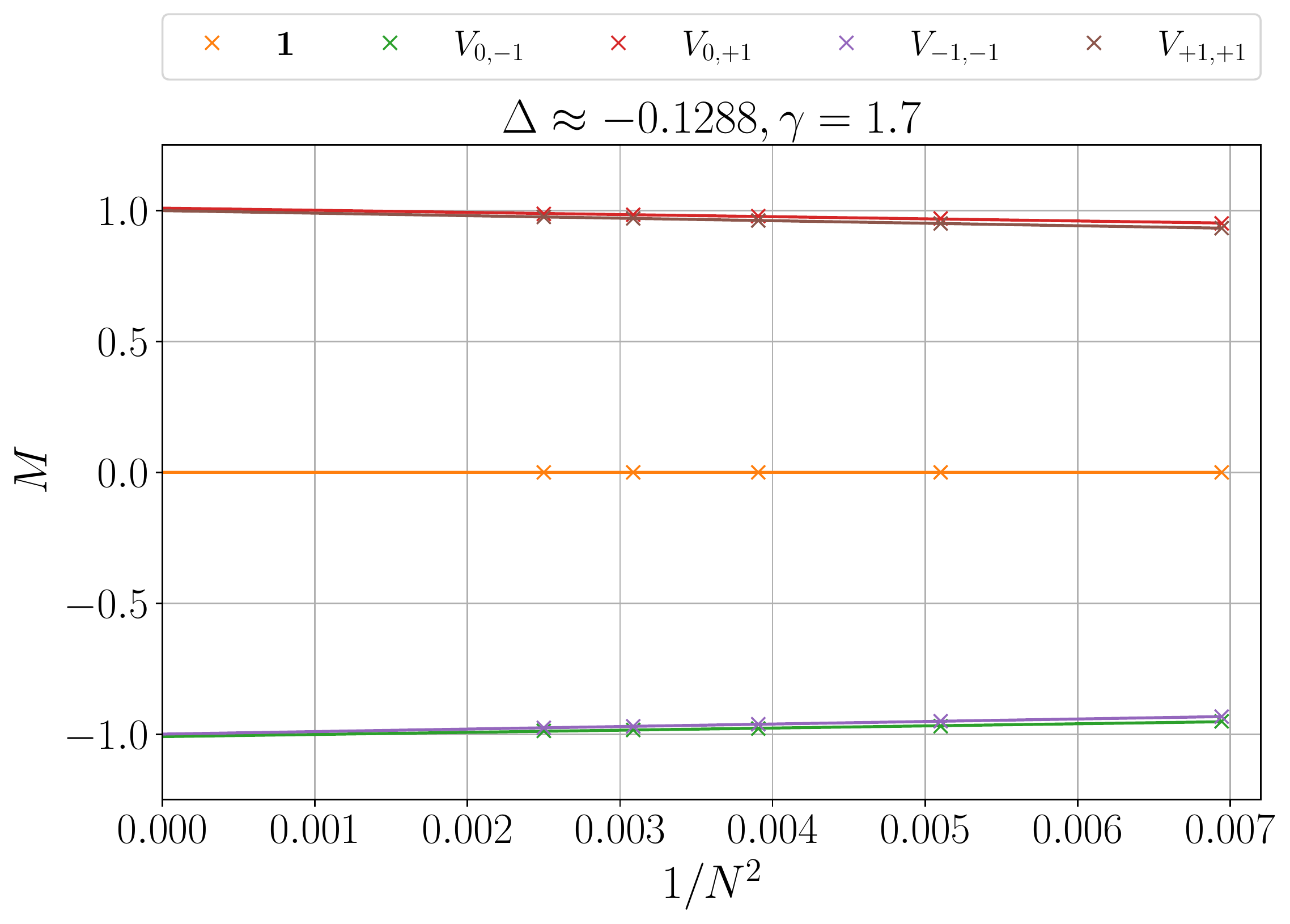}
\end{minipage}
\caption{Finite size scaling of $\mel{\psi}{M}{\psi}$ from XXZ model with $\gamma=2\pi/3$ at system sizes $N=12,14,16,18,20$.}
\label{fig:xxz_m}
\end{figure}
Our construction of lattice Kac--Moody generators works for general XXZ model. To illustrate, we study XXZ model with $\gamma=2\pi/3$ as an example of a rational CFT, and $\gamma=1.7$ as an example of an irrational CFT. 
 
As opposed to the nonabelian case, the normalization of charge density $q$ is not fixed by the Lie algebra. Thus we have freedom to add an overall normalization constant in front of $J$ and $\bar{J}$. In order to fix the normalization, we use the normalization of the CFT
 \begin{equation}
     J^{\CFT}_0 = \frac{Q}{R} + \frac{MR}{2},~~\bar{J}^{\CFT}_0 = \frac{Q}{R} - \frac{MR}{2},
 \end{equation}
 where $Q$ and $M$ are integers and $R$ is the compactification radius. We thus identify
 \begin{equation}
     \frac{R}{2}(J^{\CFT}_0 +\bar{J}^{\CFT}_0) \sim \sum_{j=1}^N {S^Z_j}
 \end{equation}
 Since they both have integer eigenvalues. Now we run the general procedure to construct the current operators on the lattice, and we obtain
 \begin{equation}
\begin{aligned}
J_n =  \sum_{j=1}^{N} e^{ijn\frac{2\pi}{N}}  \frac{1}{R}
\left[
S_{j}^{Z}+  \frac{2\gamma}{\pi \sin{\gamma}}  \left(
S_{j}^{X}S_{j+1}^{Y} - S_{j}^{Y}S_{j+1}^{X}  
\right)
\right], \\
\bar{J}_n =  \sum_{j=1}^{N}  e^{-ijn\frac{2\pi}{N}} \frac{1}{R}
\left[
{S_{j}^{Z}} -  \frac{2\gamma}{\pi \sin{\gamma}}  \left(
S_{j}^{X}S_{j+1}^{Y} - S_{j}^{Y}S_{j+1}^{X} 
\right)
\right] .
\end{aligned}
\end{equation}
They are the same as Eq.~\eqref{eq:Jnxxz} up to an overall factor. We expect that they satisfy the $U(1)$ Kac--Moody algebra at level $k=1$. This is checked and shown in Fig.~\ref{fig:xxz_k}. In order to check the emergent charge $M$, we note that in the CFT
\begin{equation}
    M = R(J^{\CFT}_0-\bar{J}^{\CFT}_0),
\end{equation}
thus on the lattice we have
\begin{equation}
    M\sim \sum_{j=1}^N \frac{4\gamma}{\pi R^2 \sin{\gamma}}  \left(
S_{j}^{X}S_{j+1}^{Y} - S_{j}^{Y}S_{j+1}^{X}  
\right).
\end{equation}
The expectation value of $M$ on low-energy eigenstates is expected to be an integer. This is indeed the case, see Fig.~\ref{fig:xxz_m}.

\end{document}